\documentclass[a4,12pt]{article}
\usepackage{subcaption}
\usepackage{multirow}
\usepackage{adjustbox}
\usepackage{caption}
\usepackage{graphicx}
\usepackage{amsmath}
\usepackage{amsfonts}
\usepackage{amssymb}
\usepackage{xcolor}
\usepackage{pdfcomment}
\usepackage{hyperref}
\usepackage[sort&compress,square,numbers,comma]{natbib}
\usepackage{mathtools}
\usepackage{float}
\usepackage{authblk}
\usepackage{hyperref}
\usepackage{enumitem}
\usepackage[pass=paperwidth=8.875in,paperheight=11in]{geometry}

\titlepage
\begin{document}
\title{Stability of the next-to-tribimaximal mixings under radiative corrections with the variation of SUSY breaking scale in MSSM}
\date{}
\author[1]{Kh. Helensana Devi \footnote{helensanakhumanthem2@manipuruniv.ac.in}}
\author[1]{K. Sashikanta Singh \footnote{ksm1skynet@gmail.com}}
\author[1,2]{N.Nimai Singh \footnote{nimai03@yahoo.com} }
\affil[1]{\small{{Department of Physics, Manipur University, Imphal-795003, India} }}
\affil[2]{Research Institute of Science and Technology, Imphal-795003, India }

\maketitle \thispagestyle{empty}
\begin{abstract}
We analyse the radiative stability of the next-to-tribimaximal mixings ($NTBM$)  with the variation of SUSY breaking scale ($m_S$) in MSSM, for both normal ordering (NO) and inverted ordering (IO) at the fixed input value of seesaw scale $M_R = 10^{15}$ GeV and two different values of $\tan \beta$. All the neutrino oscillation parameters receive varying radiative corrections irrespective of the $m_S$ values at the electroweak scale, which are all within $3\sigma$ range of the latest global fit data at low value of $\tan \beta$ (30). NO is found to be more stable than IO for all four different NTBM mixing patterns.
\end{abstract}

\titlepage
\section{Introduction} 
Neutrino oscillations have been very well established by measuring the neutrino mixing parameters $\theta_{12}$, $\theta_{23}$, $\theta_{13}$, $\Delta m^{2}_{21}$ and $\Delta m^{2}_{31}$ \cite{c2}. One of the promising candidates for
explaining the observed Pontecorvo-Maki-Nakawaga-Sakata (PMNS) mixing matrix, is the tribimaximal (TBM) \cite{tbm} which is ruled out due to the discovery of the nonzero value of $\theta_{13}$ \cite{tk,mnos,dcz,db,reno,lia,lib}. Hence, in order to accommodate $\theta_{13} \ne 0$, PMNS matrix is reproduced using the next-to-TBM (NTBM) \cite{ntbm1,ntbm2} which predicts the correlations among the phase and mixing angles. Existence of the PMNS mixing matrix, which is the analogue of the CKM matrix in the quark sector, is the consequence of diagonalisation of the neutrino mass matrix. The PMNS mixing matrix contains three mixing angles $\theta_{12}$, $\theta_{23}$, $\theta_{13}$ and a phase $\delta_{CP}$ responsible for CP violation. Two additional phases which do not influence neutrino oscillations, are added if we consider neutrinos are Majorana fermions. The measurement of a nonzero $\theta_{13}$ using reactor neutrinos in 2012, has opened the possibility to measure CP violation in the lepton sector.

The present work is a continuation of our previous work \cite{hel} on neutrino masses and mixings with varying SUSY breaking scale $m_S$ under RGEs \cite{rgg1,rgg2,rgg3,rgg4,rgg5,da,dac,hel}. We study both normal and inverted ordering neutrino mass models.  We adopt the bottom-up approach for running gauge and Yukawa couplings from low to high energy scales, and the top-down approach for running neutrino parameters from high to low energy scales, along with gauge and Yukawa couplings.

Following the discovery of a non-zero $\theta_{13}$, the originally proposed TBM (tribimaximal) mixing pattern, which initially assumed $\theta_{13}$ to be zero, was no longer considered a valid description. Consequently, extensive research efforts were directed towards exploring various TBM variants capable of accommodating a non-zero $\theta_{13}$ while accurately reproducing the PMNS (Pontecorvo-Maki-Nakagawa-Sakata) matrix. Among these variants, the next-to-TBM (NTBM) mixing scheme has gained prominence. NTBM mixing is characterized by a two-parameter family and has played a crucial role in making precise predictions and establishing correlations among the mixing angles $\theta_{ij}$ and the Dirac CP phase $\delta_{CP}$. Multiple model-independent analyses have delved into these correlations, with references available in [\cite{Kang:2014mka, Rodejohann:2011uz, He:2011gb, Albright:2008rp, Garg:2018jsg}].
It is worth emphasizing that the original TBM mixing pattern, denoted as $V_{TBM}$, explicitly predicts $\theta_{13}$ to be exactly zero. However, thorough global fit studies, as summarized in Table [\ref{tab:2}], have unequivocally demonstrated the existence of a non-zero $\theta_{13}$. Consequently, a strong incentive exists to investigate deviations from the TBM mixing pattern. Phenomenologically, small deviations from the TBM pattern can be easily parameterized by multiplication with a unitary rotation matrix. For example, the $TBM_1$ mixing scheme, given in equation \ref{eq:tbm1l}, can be seen as a modification to the TBM mixing pattern by applying a 23-rotation matrix $R_{23}$ from the right to $V_{TBM}$. In this sense, the rotation matrix $R_{23}(\kappa_1, \kappa_2)$ could also be interpreted as a perturbation to the exact TBM mixing pattern. The physical observables - namely, the three mixing angles and the Dirac CP phase $\delta_{CP}$ - are correlated via the two parameters $\kappa_1$ and $\kappa_2$, which leads us to a well-defined phenomenology.

   There are four allowed NTBM patterns \cite{ntbm} depending on the position (left or right) of the multiplication by a unitary rotation matrix to tribimaximal (TBM) mixing. The $V_{TBM}$ is given by 
\begin{equation}
V_{TBM} = 
\left({\begin{array}{ccc}
\sqrt{\frac{2}{3} }& \frac{1}{\sqrt{3}} & 0 \\ 
- \frac{1}{\sqrt{6}}  &  \frac{1}{\sqrt{3}}  &  \frac{1}{\sqrt{2}} \\ 
-  \frac{1}{\sqrt{6}}  &  \frac{1}{\sqrt{3}}  &- \frac{1}{\sqrt{2}}  \\
\end{array} }\right)
\end{equation}

The four allowed NTBM patterns are \cite{ntbm}\\
\begin{equation}{\label{eq:tbm1l}}
 U_1 = V_{TBM} R_{23},
\end{equation}

\begin{equation}{\label{eq:tbm2l}}
 U_2 = V_{TBM} R_{13}, 
\end{equation}

\begin{equation}{\label{eq:tbm2u}}
 U^2 = R_{13}V_{TBM},
\end{equation}

\begin{equation}{\label{eq:tbm3u}}
U^3 = R_{12} V_{TBM},
\end{equation}

where $R_{23}$, $R_{13}$, and $R_{12}$ are the rotation matrices defined as 

\begin{equation}
R_{23} = 
\left({\begin{array}{ccc}
1&0 & 0 \\ 
0 &  \cos \kappa_1 &  \sin  \kappa_1 e^{i\kappa_2}\\ 
0  & -\sin \kappa_1 e^{-i\kappa_2}  &\cos \kappa_1\\
\end{array} }\right),
\end{equation}

\begin{equation}
R_{13} = 
\left({\begin{array}{ccc}
\cos \kappa_1& 0 & \sin \kappa_1 e^{i\kappa_2} \\ 
0&  1 &  0\\ 
-\sin \kappa_1 e^{-i\kappa_2}  & 0  &\cos \kappa_1 \\
\end{array} }\right),
 \end{equation}

\begin{equation}
R_{12} = 
\left({\begin{array}{ccc}
\cos \kappa_1& \sin \kappa_1 e^{i\kappa_2} & 0 \\ 
- \sin \kappa_1e^{-i\kappa_2} &  \cos \kappa_1  &  0\\ 
0  & 0  &1  \\
\end{array} }\right),
 \end{equation}

where $\kappa_1$  and $\kappa_2$  are free parameters within the ranges $0\leq\kappa_1\leq\pi$ and
 $0\leq \kappa_2< 2\pi$, respectively. $U_1$, $U_2$, $U^2$, and $U^3$ are distinct NTBM mixing patterns for the $TBM_1$, $TBM_2$, $TBM^2$, and $TBM^3$ scenarios, respectively. In the present work, we calculate the values of mixing angles and $\delta_{CP}$ given by these four different mixing patterns, and these are found to be valid for certain values of $\kappa_1$  and $\kappa_2$. We check stability against radiative corrections by varying the SUSY breaking scale $m_S$ within the range of $2 - 14$ TeV, considering two different values of $\tan \beta = 30, 50$. The paper is organized as follows. NTBM is discussed in section 2. Analysis for RGEs is discussed in Section 3. Numerical analysis and results are given in section 4. Section 5 concludes the paper. 
\section{Numerical predictions in next-to-TBM (NTBM) }
NTBM is defined by multiplying $V_{TBM}$  by a unitary rotation matrix either on the left
or right. There are six possible NTBM patterns but only four patterns given in equations \ref{eq:tbm1l}-\ref{eq:tbm3u} are allowed since two  patterns: $U_P = V_{TBM} R_{12}$ and $U_P = R_{23} V_{TBM}$ are already excluded as they predict zero $\theta_{13}$. The four NTBM patterns provide formulas for the mixing angles and $\delta_{CP}$ in terms of free parameters $\kappa_1$ and $\kappa_2$, which characterize the two-parameter family of NTBM patterns \cite{ntbm}. The four patterns are given below:
\subsection{$TBM_1$ pattern}

$ \displaystyle \tan \theta_{23} = \biggl|\frac{3\sqrt{2}\cos \kappa_1+ 2\sqrt{3}e^{i \kappa_2}\sin \kappa_1}{3\sqrt{2}\cos \kappa_1-
2\sqrt{3}e^{i \kappa_2}\sin \kappa_1 }\biggl|$, 
$\displaystyle  {\tan \theta_{12}} = \frac{|\cos \kappa_1| }{\sqrt{2}}$, 
$ \displaystyle \sin \theta_{13} = \frac{|\sin \kappa_1| }{\sqrt{3}}$ and 
\begin{equation}{\label{eq:nt}}
\displaystyle  \sin \delta_{CP} =\frac{ (\sin2 \kappa_1)\sin \kappa_2 (\cos 2 \kappa_1 + 5)}{[ (\cos \kappa_1 + 5)^2 - ( 2\sqrt{6} \sin 2 \kappa_1 \cos \kappa_2)^2 ]^{1/2}}
\end{equation}

\subsection{$TBM_2$ pattern}

$ \displaystyle \tan \theta_{23} = \biggl|\frac{\sqrt{3}e^{i \kappa_2}\sin \kappa_1 - 3\cos \kappa_1}{3\cos \kappa_1 +
\sqrt{3}e^{i \kappa_2}\sin \kappa_1}\biggl|$, 
$\displaystyle  {\tan \theta_{12}} = \frac{1 }{\sqrt{2}|\cos\kappa_1|}$, 
$ \displaystyle \sin \theta_{13} = \sqrt{\frac{2}{3}}|\sin \kappa_1|$ and 
\begin{equation}{\label{eq:nt1}}
\displaystyle  \sin \delta_{CP} =\frac{ (\sin2\kappa_1)(\cos 2 \kappa_1 + 2)\sin \kappa_2}{[ (2\cos^2 \kappa_1  + 1)^2 - 3 \sin  ^2 2\kappa_1 \cos ^2 \kappa_2 ]^{1/2}}
\end{equation}
\subsection{$TBM^2$ pattern}

$ \displaystyle \tan \theta_{23} = \frac{1}{|\cos\kappa_1|}$, 
$\displaystyle  \frac{\tan \theta_{12}}{\sqrt{2}} = \biggl|\frac{\cos \kappa_1 + e^{i\kappa_2}\sin \kappa_1 }{2\cos \kappa_1 - e^{i \kappa_2}\sin \kappa_1}\biggl|$, 
$ \displaystyle \sin \theta_{13} = \frac{|\sin \kappa_1|}{\sqrt{2}}$ and 
\begin{equation}{\label{eq:nt2}}
\displaystyle  \sin \delta_{CP} =\frac{(\sin2\kappa_1)(\cos 2 \kappa_1+ 3)\sin \kappa_2}{2\biggl[ \biggl(-2 \sin 2\kappa_1\cos \kappa_2 + 3 \cos^2\kappa_1 + 1\biggl)\biggl( 1 + \sin 2 \kappa_1 \cos \kappa_2\biggl)\biggl]^{1/2}}
\end{equation}

\subsection{$TBM^3$ pattern}
$ \displaystyle \tan \theta_{23} = |\cos\kappa_1|$, 
$\displaystyle  \frac{\tan \theta_{12}}{\sqrt{2}} = \biggl|\frac{\cos \kappa_1 + e^{i \kappa_2}\sin \kappa_1}{2\cos \kappa_1 - e^{i \kappa_2}\sin \kappa_1}\biggl|$, 
$ \displaystyle \sin \theta_{13} = \frac{|\sin\kappa_1|}{\sqrt{2}}$ and 
\begin{equation}{\label{eq:nt3}}
\displaystyle  \sin \delta_{CP} =-\frac{(\sin2\kappa_1)(\cos 2 \kappa_1 + 3)\sin \kappa_2}{2\biggl[ \biggl(-2 \sin 2\kappa_1\cos \kappa_2+ 3 \cos^2\kappa_1+ 1\biggl)\biggl( 1 + \sin 2  \kappa_1\cos \kappa_2 \biggl)\biggl]^{1/2}}
\end{equation}

\begin{figure}[H]
\centering
\begin{subfigure}{.62\textwidth}
    \includegraphics[scale=0.48]{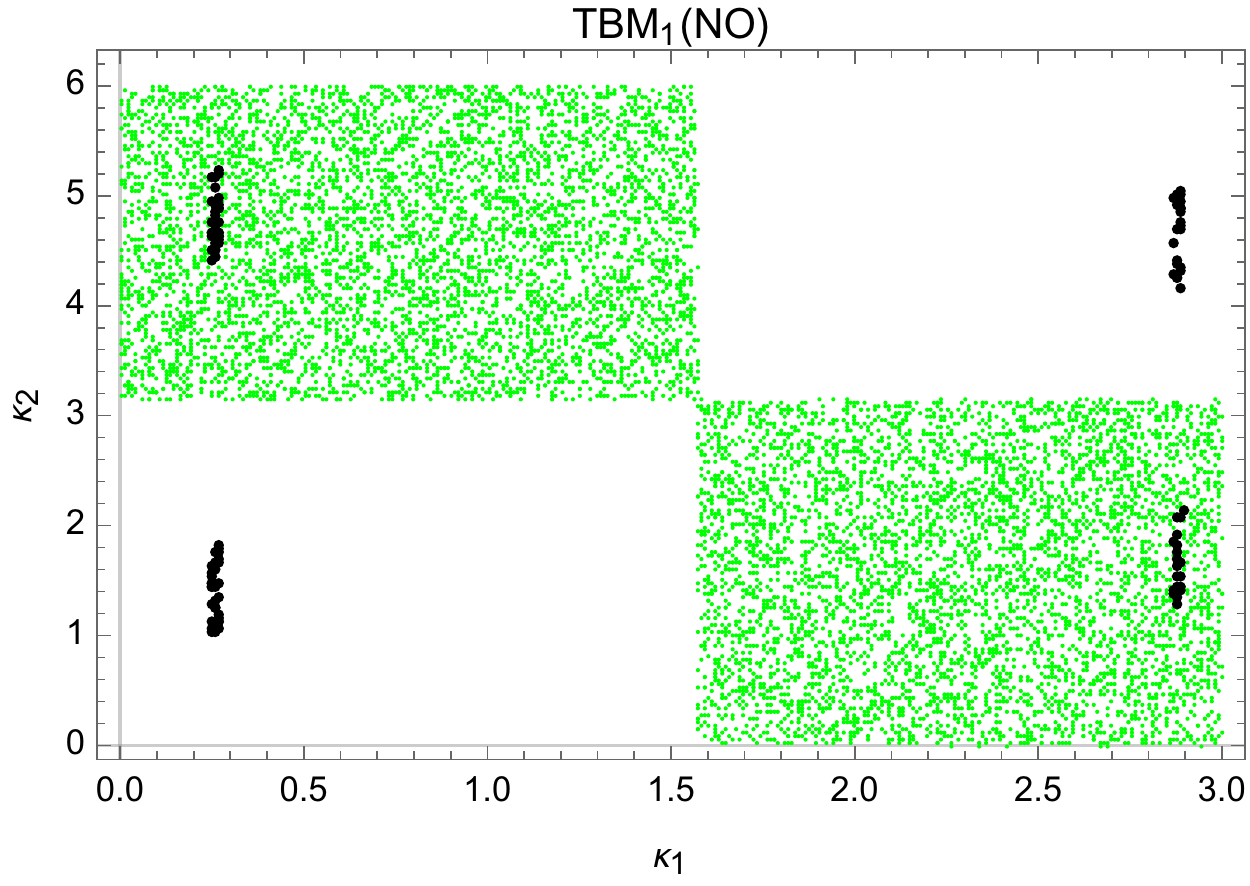}
\end{subfigure}
\begin{subfigure}{0.23\textwidth}
   \includegraphics[scale=0.48]{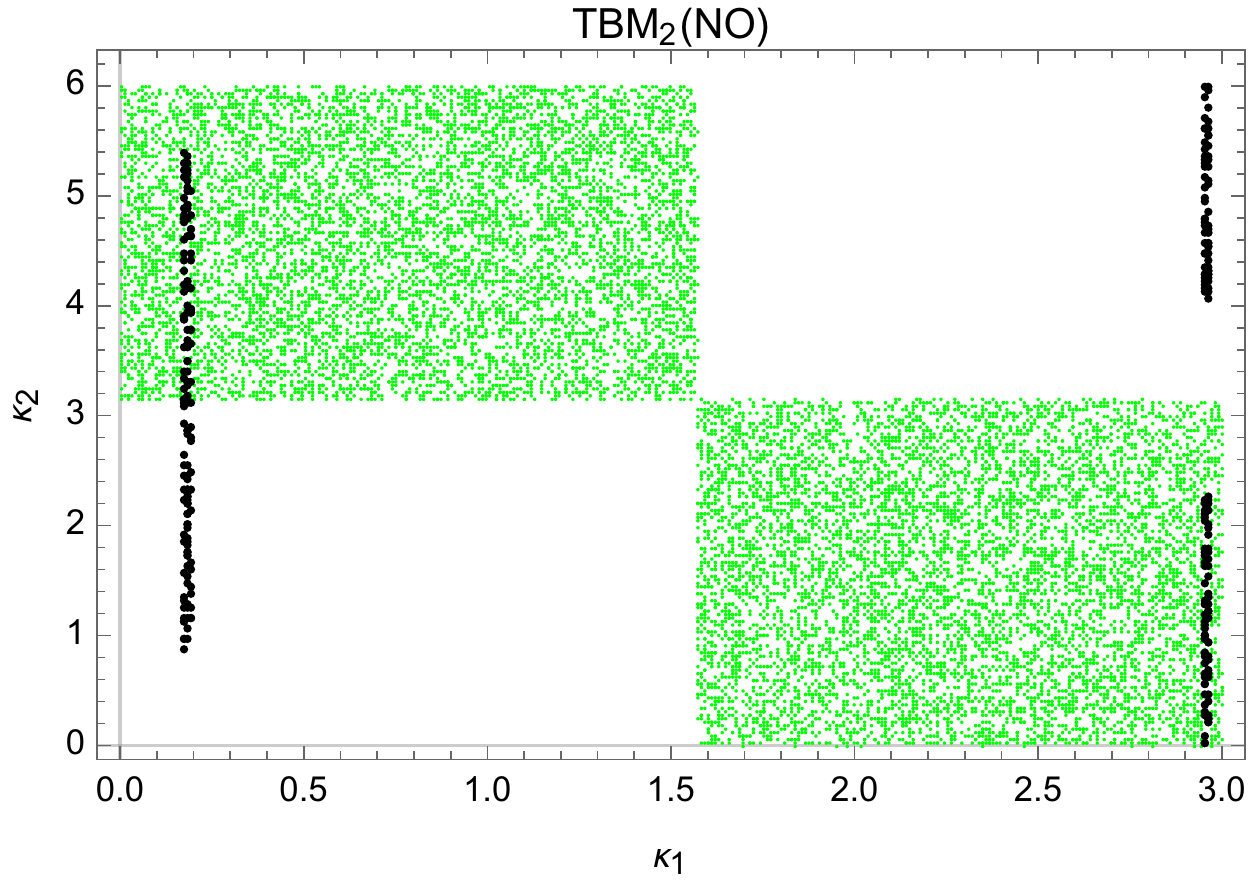}
\end{subfigure}
\begin{subfigure}{0.62\textwidth}
  \includegraphics[scale=0.48]{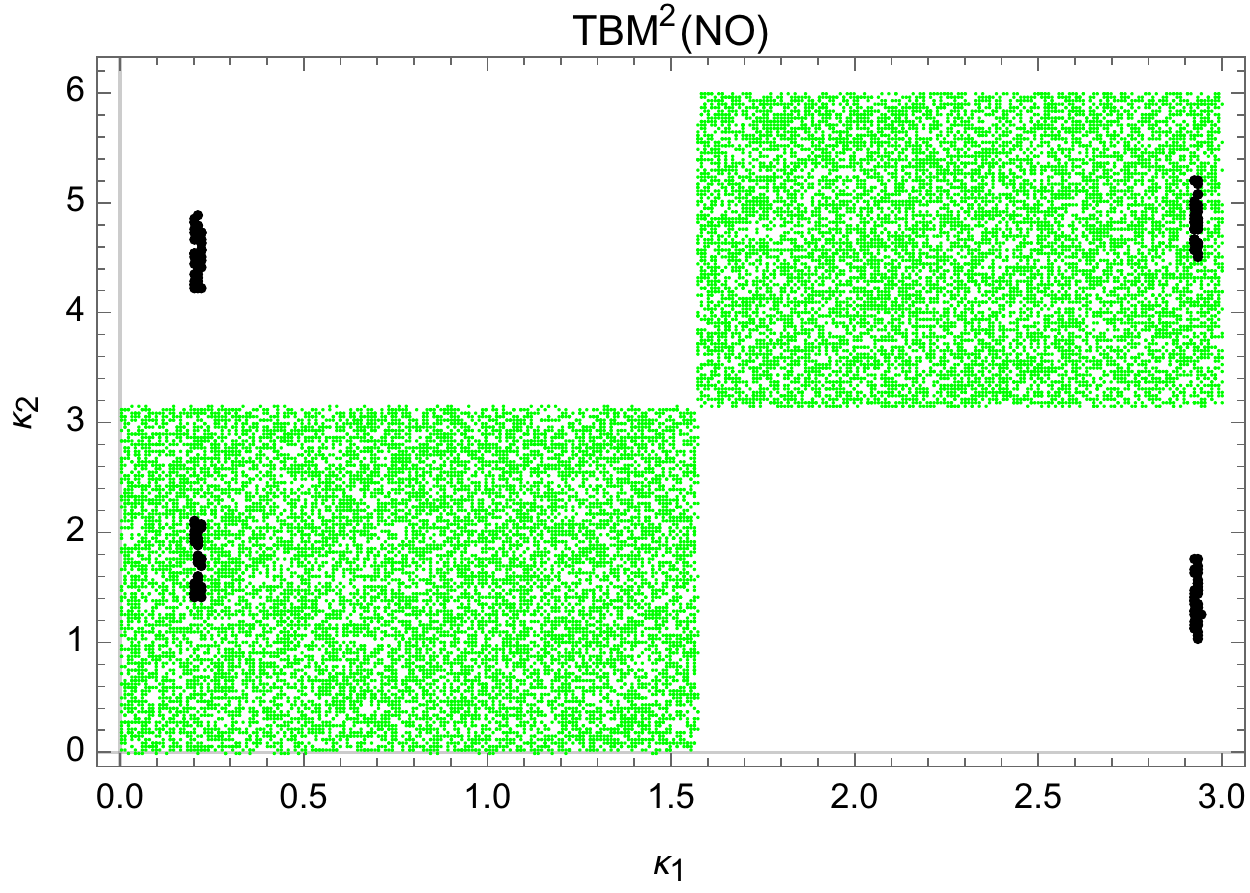}
\end{subfigure}
\begin{subfigure}{0.23\textwidth}
  \includegraphics[scale=0.48]{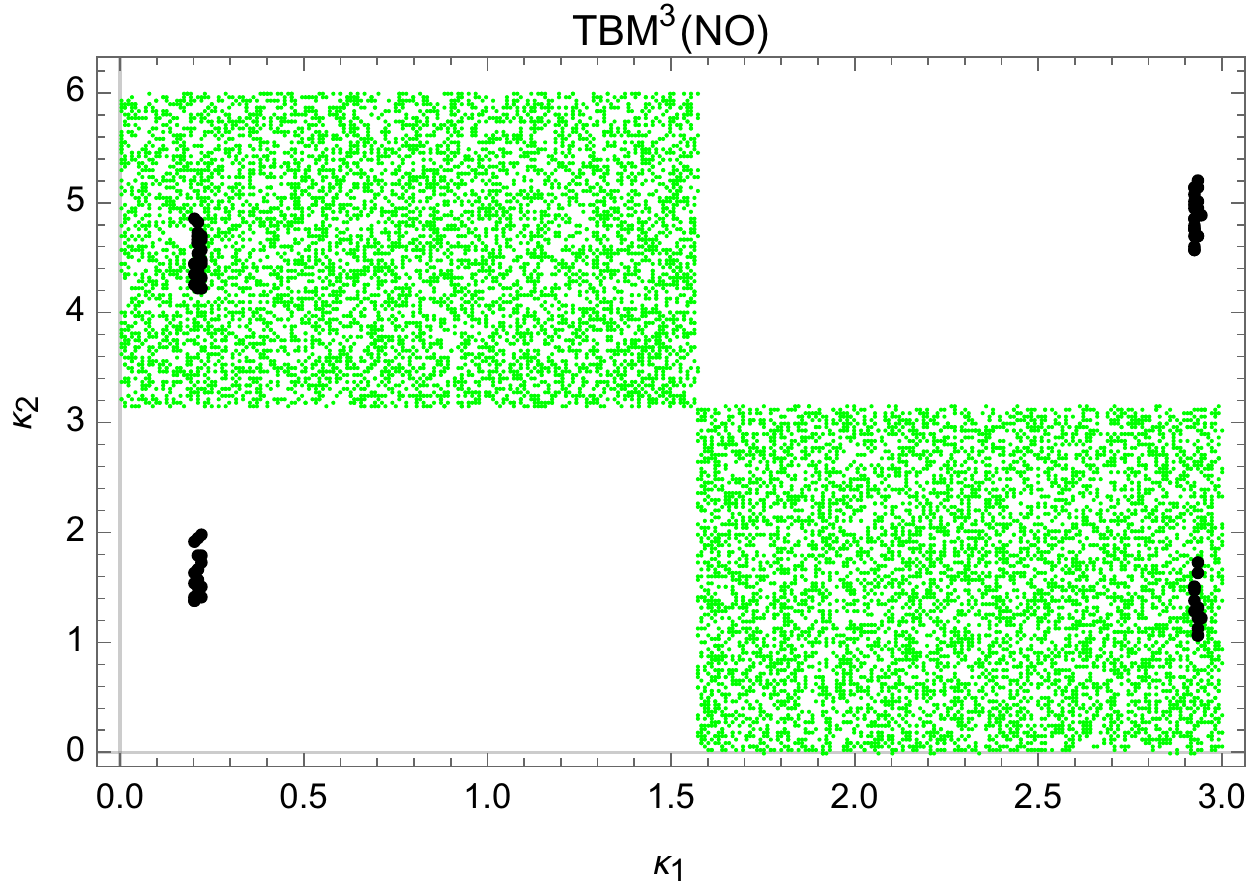}
\end{subfigure}
\caption{\footnotesize{Constraints on the two free parameters $\kappa_1$ and $\kappa_2$ are applied in the context of the NO spectrum. The $3\sigma$-allowed mixing angles are shaded in black, and regions where $\sin\delta_{CP} > 0$ are shaded in green, both for the four different mixing patterns of NTBM.}}
\label{fig:figab}

\end{figure}

\begin{figure}[H]
\centering
\begin{subfigure}{.62\textwidth}
    \includegraphics[scale=0.48]{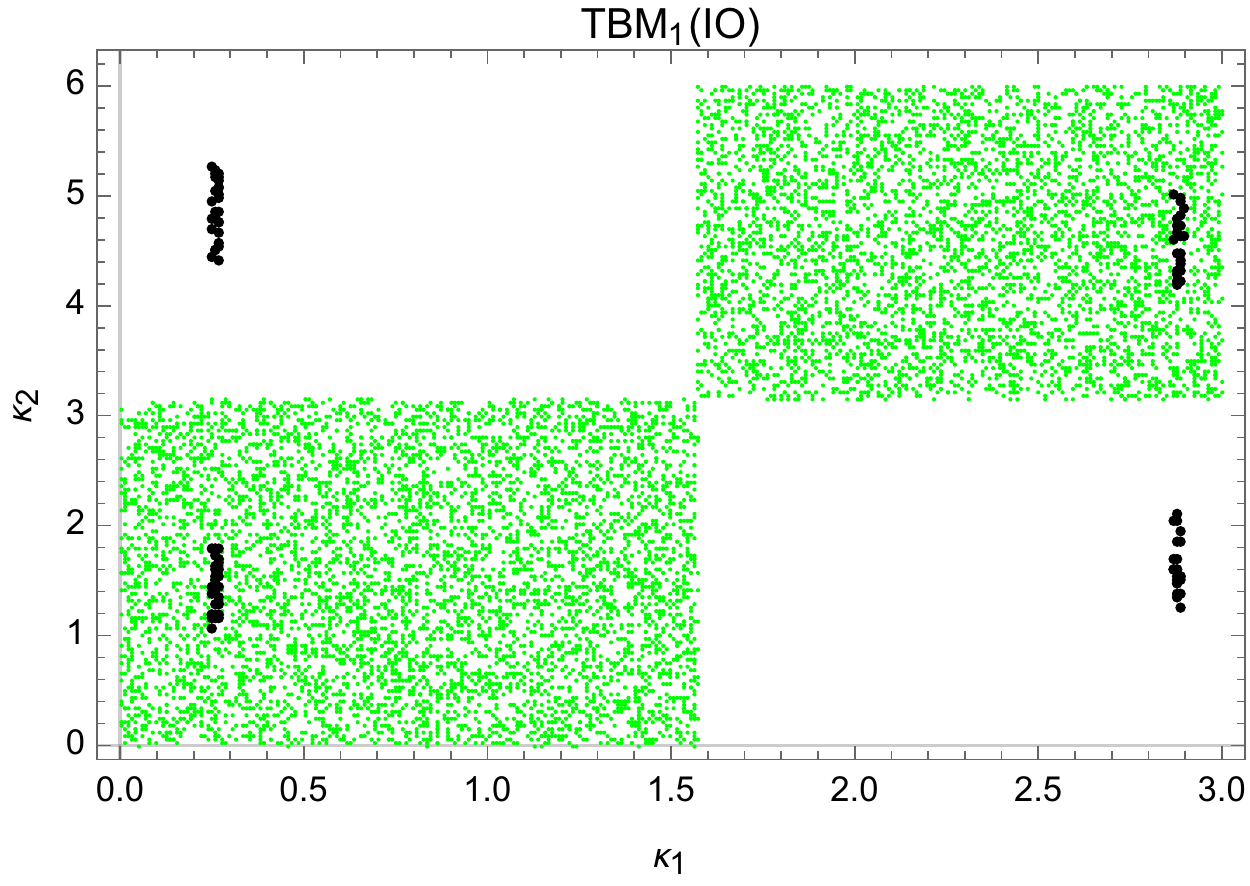}
\end{subfigure}
\begin{subfigure}{0.23\textwidth}
   \includegraphics[scale=0.48]{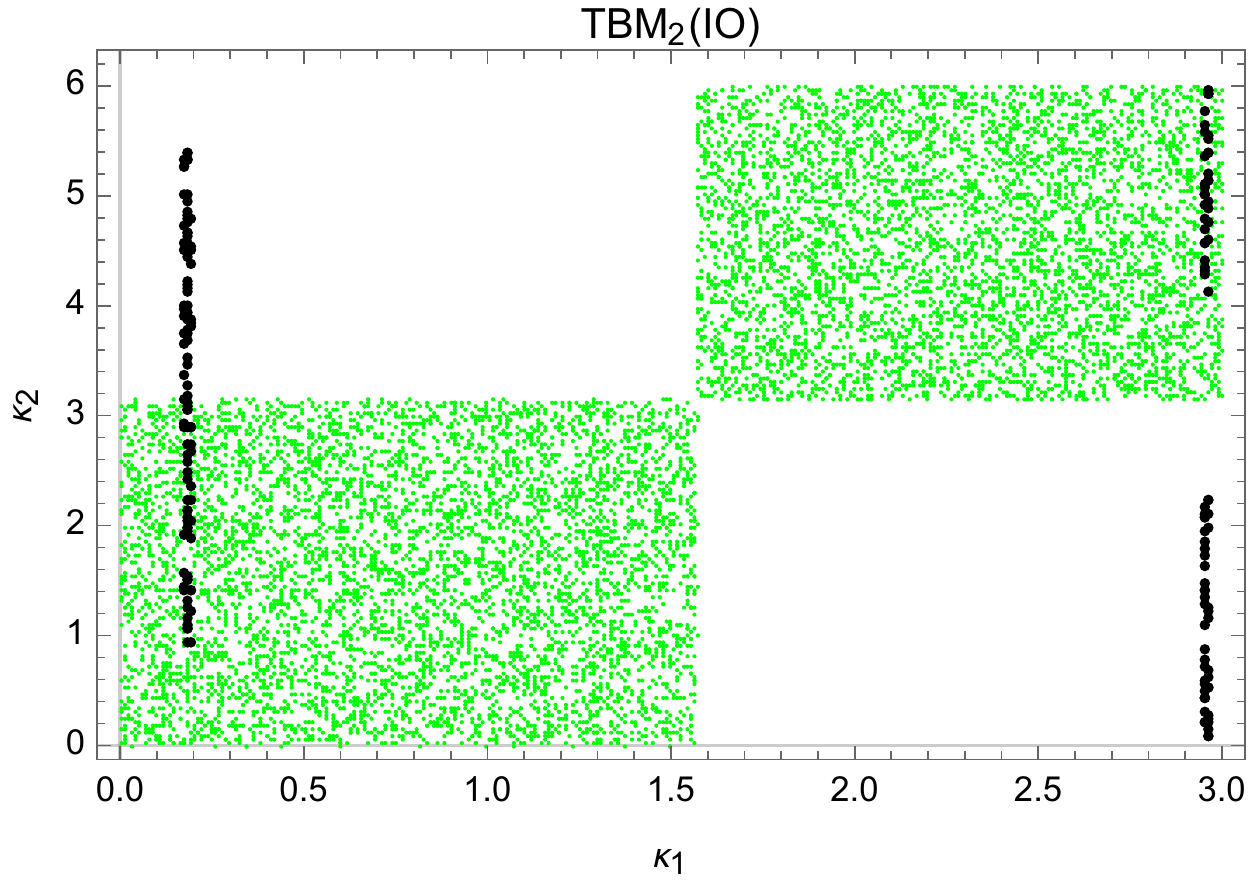}
\end{subfigure}
\begin{subfigure}{0.62\textwidth}
  \includegraphics[scale=0.48]{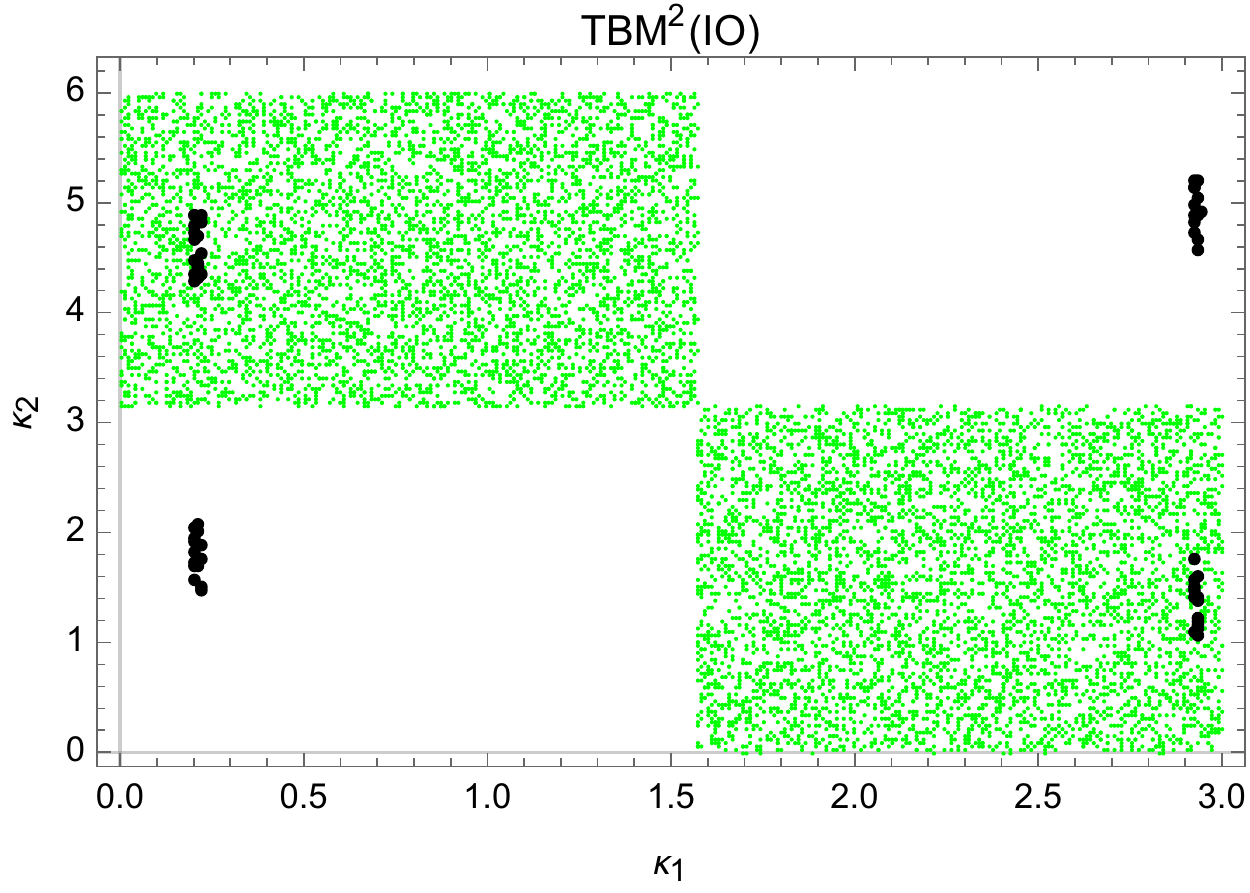}
\end{subfigure}
\begin{subfigure}{0.23\textwidth}
  \includegraphics[scale=0.48]{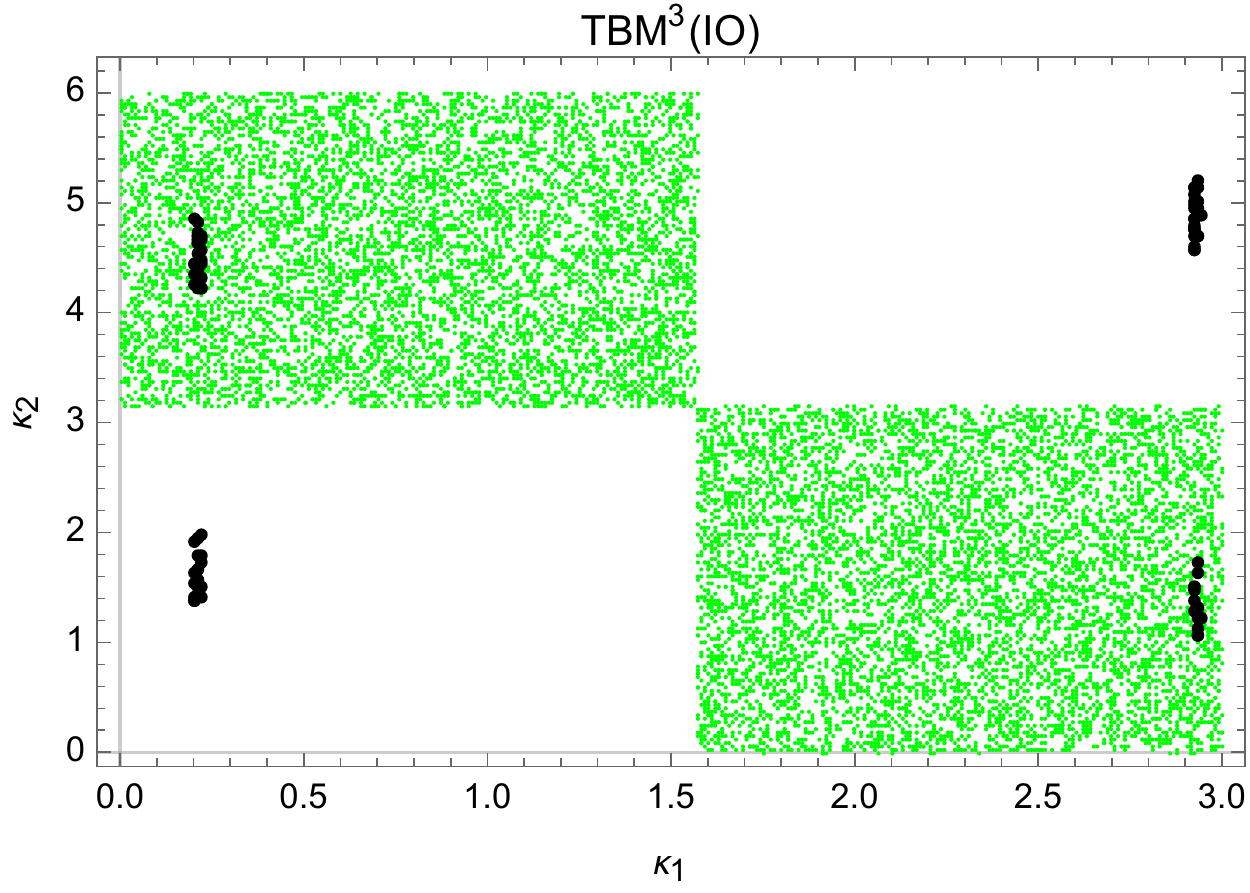}
\end{subfigure}
\caption{\footnotesize{Constraints on the two free parameters $\kappa_1$ and $\kappa_2$ are imposed within the inverted ordering (IO) spectrum. The $3\sigma$-allowed mixing angles are shaded in black, while regions where $\sin\delta_{CP} < 0$ are shaded in green, for the four different mixing patterns of NTBM.}}
\label{fig:figac}

\end{figure}

 We impose the conditions $\sin \delta_{CP} > 0$ (NO) and $\sin \delta_{CP} < 0$ (IO) to constrain the two free parameters $\kappa_1$ and $\kappa_2$ for all NTBM scenarios. The best-estimated numerical values of $\theta_{ij}$ and $\delta_{CP}$ for $TBM_1$, $TBM_2$, $TBM^2$, and $TBM^3$ are provided in Table \ref{tab:i3} for specific choices of $\kappa_1$ and $\kappa_2$ in each case, both for NO and IO. We use values of $\kappa_1$ and $\kappa_2$ that fall within the allowed regions depicted in Figures \ref{fig:figab} and \ref{fig:figac} for both NO and IO, respectively

\section{Analysis  for RGEs}
Numerical analysis of Renormalization Group Equations (RGEs) \cite{da,rg1,beta} is conducted in two successive steps: first, bottom-up running; and second, top-down running. Two-loop Renormalization Group Equations (RGEs) for gauge and Yukawa couplings are provided in Appendix A for both the Standard Model (SM) and the Minimal Supersymmetric Standard Model (MSSM). The RGEs for neutrino oscillation parameters are presented in Appendix B.
 
\subsection{Bottom-up Running}

 Bottom-up running is used to extract the values of gauge and Yukawa couplings at high energy scale using RGEs which can be divided into three regions, $ m_Z < \mu < m_t$, $m_t < \mu <m_S$, and $m_S< \mu <M_R$. We use recent experimental data \cite{c2,pgd} as initial input values at low energy scale,  which are given in Table \ref{tab:1z}. 
 
\begin{table}[H]
\begin{adjustbox}{width=0.99\textwidth}
\centering
\begin{tabular}{lll}
\hline
Mass(GeV)& Coupling constants & Weinberg mixing angle \\ 
\hline
$m_Z(m_Z)$= 91.1876& $\alpha_{em}^{-1}(m_Z)$ =127.952 $\pm$ 0.009 & $\sin^2\theta_W(m_Z)$ = 0.23121 $\pm 0.00017$\\
 $m_t(m_t)$= 172.76 & $\alpha_{s}(m_Z)$ =0.1179$\pm$ 0.009\\
 $m_b(m_b)$= 4.18 &&\\
 $m_{\tau}(m_{\tau})$=1.77&&\\
\hline 
\end{tabular} 
\end{adjustbox}
\caption{ \footnotesize{Current experimental data for fermion masses, gauge coupling constants and Weinberg mixing angle.}}
\label{tab:1z}
\end{table}
 
 We calculate the values of gauge couplings, $\alpha_{2}$  for $SU(2)_{L}$ and $\alpha_{1}$ for $U(1)_{Y}$, by using  $\displaystyle \sin^2 \theta_{W}(m_{Z}) = \alpha_{em}(m_Z)/\alpha_2(m_Z)$ and matching condition, 
\begin{equation}
\frac{1}{\alpha_{em}(m_Z)} =\frac{5}{3}\frac{1}{\alpha_1(m_Z)} + \frac{1}{\alpha_2(m_Z)}.
\label{b}
\end{equation}
The normalised couplings \cite{da}, $g_i = \sqrt{4\pi \alpha_i}$, where  $\alpha_i$'s are the gauge couplings and $i=1, 2, 3$ denote electromagnetic, weak and strong couplings respectively. One-loop gauge couplings RGEs  \cite{bb1} for the evolution from $m_Z$ scale to $m_t$ scale, are given below:
\begin{equation}
\frac{1}{\alpha_i(\mu)} = \frac{1}{\alpha_{i}(m_Z)} - \frac{b
_i}{2\pi}ln\frac{\mu}{m_Z},
\label{d}
\end{equation}
where $m_Z\leq \mu \leq m_t$  and $b_i$ = (5.30, -0.50, -4.00) for non-SUSY case. Using QED-QCD rescaling factor $\eta$ \cite{qcd}, fermion masses at $m_t$ scale, are given by $\displaystyle m_b(m_t) = m_b(m_b)/\eta_b$ and $\displaystyle  m_{\tau}(m_t) = m_{\tau}(m_\tau)/\eta_{\tau}$, where $\displaystyle  \eta_{b} = 1.53$ and $\eta_{\tau} = 1.015$. The Yukawa couplings at $m_t$ scale are given by $\displaystyle  h_t(m_t) = m_t(m_t)/v_0$, $\displaystyle  h_b(m_t) = m_b(m_b)/v_0 \eta_b$, and $\displaystyle  h_\tau(m_\tau) = m_\tau(m_\tau)/v_0\eta_\tau$, where $v_0$ = 174 GeV is the vacuum expectation value (VEV) of SM Higgs field. The calculated numerical values for fermion masses, Yukawa and gauge couplings at $m_t$ scale, are given in Table \ref{tab:1b}.

\begin{table}[H]
\centering
\begin{tabular}{ccc} 
\hline
Fermions masses & Yukawa Couplings & Gauge Couplings \\
\hline
$m_t(m_t)$ = 172.76 GeV& $h_t(m_t)$ = 0.9928&$g_1(m_t)$ =  0.4635\\
$m_b(m_t)$ = 2.73 GeV & $h_b(m_t)$ = 0.0157&$g_2(m_t)$ = 0.6511\\
$m_{\tau}(m_t)$ = 1.75 GeV  & $h_{\tau}(m_t)$ = 0.0100&$g_3(m_t)$ = 1.1890\\
\hline
\end{tabular} 
\caption{\footnotesize{ Numerical calculated values for fermion masses, Yukawa  and gauge couplings at $m_t$ scale.}}
\label{tab:1b}
\end{table}
 
The evolution of gauge and Yukawa couplings for running from $m_t$ to the $M_R$ scale using RGEs, are given in Appendix A. The following matching conditions are applied at the transition point from SM ($m_t<\mu<m_S$) to MSSM ($m_S<\mu<M_R$)  at $m_S$ scale, as

\begin{equation}
\left.
\begin{array}{l}
g_i(SUSY) = g_i(SM) \\
    h_t(SUSY) = \frac{h_t(SM)}{\sin\beta}= h_t(SM) \times \frac{\sqrt{1+ \tan^2\beta}}{\tan\beta}\\
h_b(SUSY) = \frac{h_b(SM)}{\cos\beta}= h_b(SM) \times \sqrt{1+ \tan^2\beta}\\
h_\tau(SUSY) = \frac{h_\tau(SM)}{\cos\beta}= h_\tau(SM) \times \sqrt{1+ \tan^2\beta} 
\end{array}
\right\}
\label{eq:t1}
\end{equation}
   
    In the present work, we have observed the following trends at input values of $\tan\beta = 30$ and $\tan\beta = 50$. At $\tan\beta = 30$, both $h_t$ and $h_b$ decrease as the $m_S$ scale increases, while $h_{\tau}$ increases with the increment in the $m_S$ scale due to its dependence on $\tan\beta$, as demonstrated in Eq. (\ref{eq:t1}). These trends are illustrated in Table \ref{tab:1c}, which will serve as input values for the subsequent top-down running at the high energy scale $M_R$. On the other hand, at $\tan\beta = 50$, all gauge and Yukawa couplings decrease with increasing $m_S$, as shown in Table \ref{tab:1d}.

 \begin{table}[H]
\centering
\begin{tabular}{cccc|ccc}
\hline 
$m_S$(TeV) &$h_t$ &$h_b$ &$h_\tau$ &$g_1$ &$g_2$& $g_3$\\
\hline
2& 0.5884& 0.1917&   0.2449 &    0.6591&    0.7024 &  0.7357 \\ 
4&  0.5753&   0.1912&   0.2471 &0.6548&      0.6980&  0.7328 \\ 
6 & 0.5662&    0.1908&   0.2488&     0.6514&    0.6945&  0.7305  \\ 
8&   0.5621&    0.1907&    0.2496&   0.6498&    0.6927&  0.7293 \\ 
10 &   0.5582&0.1906&    0.2505&     0.6481&   0.6910&  0.7282 \\  
12 &  0.5545&   0.1904&   0.2513&   0.6464&   0.6893&  0.7271\\
14 & 0.5528&   0.1903&  0.2517&    0.6456&   0.6885&  0.7265 \\
\hline 
\end{tabular} 
\caption{ \footnotesize{Yukawa and gauge couplings were evaluated at $M_R = 10^{15}$ GeV for $\tan\beta = 30$, considering different choices of the $m_S$ scale.}}
\label{tab:1c}
\end{table}

 \begin{table}[H]
\centering
\begin{tabular}{cccc|ccc}
\hline 
$m_S$(TeV) &$h_t$ &$h_b$ &$h_\tau$ &$g_1$ &$g_2$& $g_3$\\
\hline
2&  0.6262&  0.4097&   0.5189&     0.6587&     0.7017 &   0.7353 \\ 
4& 0.6077&     0.4012&  0.5171 &  0.6544 &     0.6973&  0.7324 \\ 
6 & 0.5951&    0.3954&   0.5160&     0.6511&    0.6939&  0.7301  \\ 
8&   0.5895&    0.3927&    0.5155&   0.6494&    0.6922&  0.7290 \\ 
10 &   0.5842&0.3902&    0.5151&     0.6477&   0.6905&  0.7279 \\  
12 &  0.5793&   0.3878&   0.5147&   0.6461&   0.6888&  0.7268\\
14 & 0.5689&   0.3819&  0.5086&    0.6453&   0.7034&  0.7263 \\
\hline 
\end{tabular} 
\caption{ \footnotesize{Yukawa and gauge couplings were evaluated at $M_R = 10^{15}$ GeV for $\tan\beta = 50$, considering different choices of the $m_S$ scale.}}
\label{tab:1d}
\end{table}

\subsection{Top-down Running }
We use top-down Running approach to study the stability for four patterns using RGEs against varying $m_S$  at fixed value of seesaw scale $M_R = 10^{15} GeV$ and $\tan \beta = 30,50$. In this running, we use the values of Yukawa and guage couplings which were earlier estimated at $M_R$ scale, as initial inputs. We consider simple mass relations for both NO and IO in order to minimize the number of input free parameters and both their sums of the three neutrino mass eigenvalues, $|\Sigma m_i|$ are all within the favorable range given by latest cosmological bound \cite{g,h}. We take the two Majorana phases $\psi_1$ and $\psi_2$ to be 0 and 180 respectively. We constraint the value of Dirac CP phase $\delta_{CP}$ at $180^0$. Using all the necessary mathematical frameworks, we analyze the radiative nature of neutrino parameters like neutrino masses, mixings, CP phases, using top-down approach  with the variations of $m_S$ scale at fixed value of $\tan\beta$ for all allowed mixing pattern. In this work we study the stability of $TBM_1$, $TBM_2$, $TBM^2$ and $TBM^3$ considering the current experimental data given in Table \ref{tab:2} . We assume a relation among the three neutrino mass eigenstates for all the four NTBM mixing patterns. The value of the free parameters $\kappa_1$ and $\kappa_2$ are considered which satisfy the condition $\sin \delta_{CP} > 0$ for NO and  $\sin \delta_{CP} < 0$ for IO. The input set at high energy scale is given in Table \ref{tab:i3}. 

\begin{table}[H]
\centering
\begin{adjustbox}{width=1.05\textwidth}
\begin{tabular}{cccc}

\hline Parameters & Best-fit  & 2$\sigma$& 3$\sigma$\\
\hline
$\Delta m^2_{21} [10^{-5} eV^2]$ &$ 7.50^{+0.22}_{-0.20}$  & 7.12 - 7.93& 6.94 - 8.14  \\
$\Delta m^2_{31} [10^{-3} eV^2]$ (NO) & $ 2.55^{+0.02}_{-0.03}$   & 2.49 - 2.60 & 2.47 - 2.63\\
$\Delta m^2_{31} [10^{-3} eV^2]$ (IO) & $ 2.45^{+0.02}_{-0.03}$   & 2.39 - 2.50 & 2.37 - 2.53\\
$\theta_{23}(NO)$ & 49.26 $ \pm$ 0.79 & 47.37 - 50.71 & 41.20 - 51.33\\
$\theta_{23}(IO)$ & $49.46 ^{+0.60}_{-0.97}$ & 47.35 - 50.67 & 41.16 - 51.25\\
$\theta_{12}$ & $ 34.3 \pm 1.0$&  32.3 - 36.4&  31.4 - 37.4 \\
$\theta_{13}(NO)$ & $ 8.53^{+0.13}_{-0.12}$ & 8.27 - 8.79 &8.13 - 8.92 \\
$\theta_{13}(IO)$ & $ 8.58^{+0.12}_{-0.14}$ & 8.30 - 8.83 &8.17 - 8.96 \\
$\delta_{CP} (NO)$ &$ 194^{+24}_{-22}$ &  152 - 255 & 128 - 359\\
$\delta_{CP} (IO)$ &$ 284^{+26}_{-28}$ &  226 - 332 & 200 - 353\\
$|\Sigma m_i|$ &  $<$ 0.12 eV ; $\ge$ 0.06 eV &&\\
\hline 
\end{tabular} 
\end{adjustbox}
\caption{ \footnotesize{Current best-fit values, 1$\sigma$ errors, and 2$\sigma$ and 3$\sigma$ intervals for the neutrino oscillation parameters from global data which are adopted from References \cite{de,c2,pgd}.}}
\label{tab:2}
\end{table}

\begin{table}[H]
\centering
\begin{adjustbox}{width=0.8\textwidth}
\centering
\begin{tabular}{llllll}
\hline 
$\nu$&& $TBM_1$&$TBM_2$&$TBM^2$&$TBM^3$\\
input&&$\kappa_1 = 2.8916$&$\kappa_1 = 2.958$&$\kappa_1 = 2.931593$&$\kappa_1=2.93$\\
parameters&&$\kappa_2=  1.641592$&$\kappa_2= 1.483182$&$\kappa_2= 4.92477$&$\kappa_2= 1.2832$\\
\hline
$m_1$(eV)&&0.012&0.012&0.012&0.012\\ 
$m_2$(eV)&& 1.245 $\times m_1$&1.245 $\times m_1$&1.245 $\times m_1$&1.245 $\times m_1$\\ 
$m_3$(eV)&& 4.392$\times m_1$&4.392$\times m_1$&4.392$\times m_1$&4.392$\times m_1$\\ 
$\theta_{12}/^0$&NO&34.416&35.7234&33.9229&33.2824\\ 
$\theta_{23}/^0$&& 45.8099 &45.5314&45.6363&44.3539 \\ 
$\theta_{13}/^0$ & &8.212 &8.572&8.476&8.540\\  
$\psi_1/^0$&&0&0&0&0 \\
$\psi_2/^0$ &&180&180&180&180\\
$\delta/^0$ &&208.581&200.957&203.939&203.864\\
\hline 
$\nu$&& $TBM_1$&$TBM_2$&$TBM^2$&$TBM^3$\\
input&&$\kappa_1 = 0.25$&$\kappa_1 = 2.985$&$\kappa_1 = 0.21$&$\kappa_1=2.93$\\
parameters&&$\kappa_2= 1.5$&$\kappa_2= 4.8$&$\kappa_2= 4.5$&$\kappa_2= 5$\\
\hline

$m_3$(eV)&& 0.009 & 0.009& 0.009& 0.009\\ 
$m_1$(eV)&& 5.661 $\times m_3$&5.661 $\times m_3$&5.661 $\times m_3$&5.661 $\times m_3$\\ 
$m_2$(eV)&& 5.696$\times m_3$&5.696$\times m_3$&5.696$\times m_3$&5.696$\times m_3$\\ 
$\theta_{12}/^0$&IO&34.416&35.7234&33.9229&33.2824\\ 
$\theta_{23}/^0$&& 45.8099 &45.5314&45.6363&44.3539 \\ 
$\theta_{13}/^0$ & &8.212 &8.572&8.476&8.540\\  
$\psi_1/^0$&&0&0&0&0 \\
$\psi_2/^0$ &&180&180&180&180\\
$\delta/^0$ &&208.582&200.957&203.939&203.87\\
\hline 
\end{tabular} 
\end{adjustbox}
\caption{ \footnotesize{The allowed input set of neutrino parameters at the high-energy scale $M_R = 10^{15}$ GeV and $\tan \beta = 30, 50$ encompasses four different patterns. $\theta_{23}$, $\theta_{12}$, $\theta_{13}$, and $\delta_{CP}$ are estimated values, while the others are arbitrary allowed input values. In this paper, we consider both the NO and IO.}}
\label{tab:i3}
\end{table}

\section{Numerical Analysis  and Results}
\begin{figure}[H]
\centering
\begin{subfigure}{0.54\textwidth}
    \includegraphics[scale=0.78]{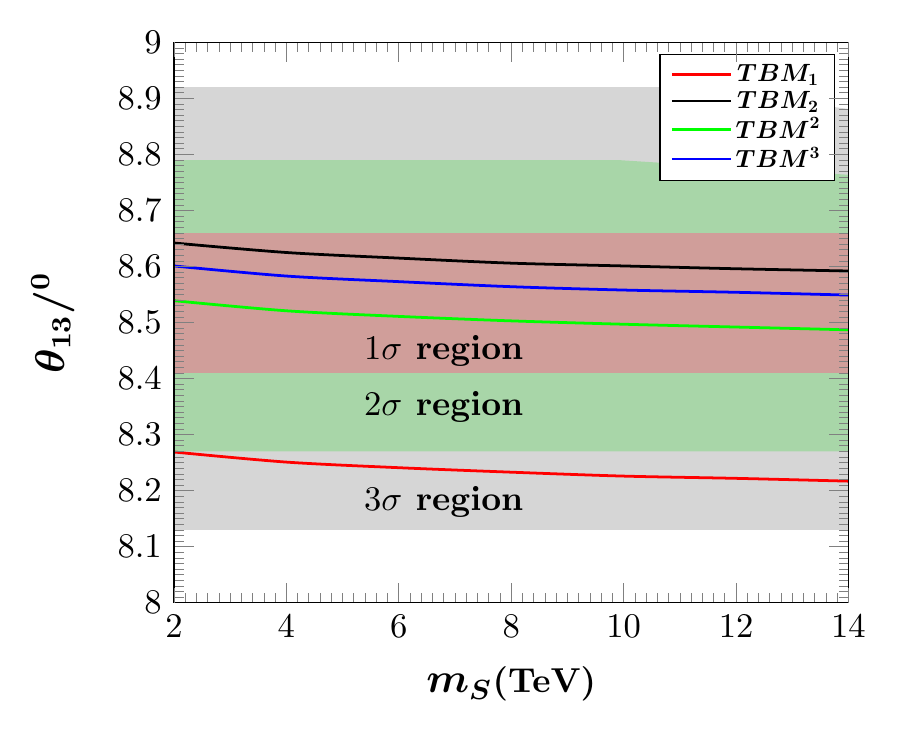}
\end{subfigure}
\begin{subfigure}{0.43\textwidth}
    \includegraphics[scale=0.78]{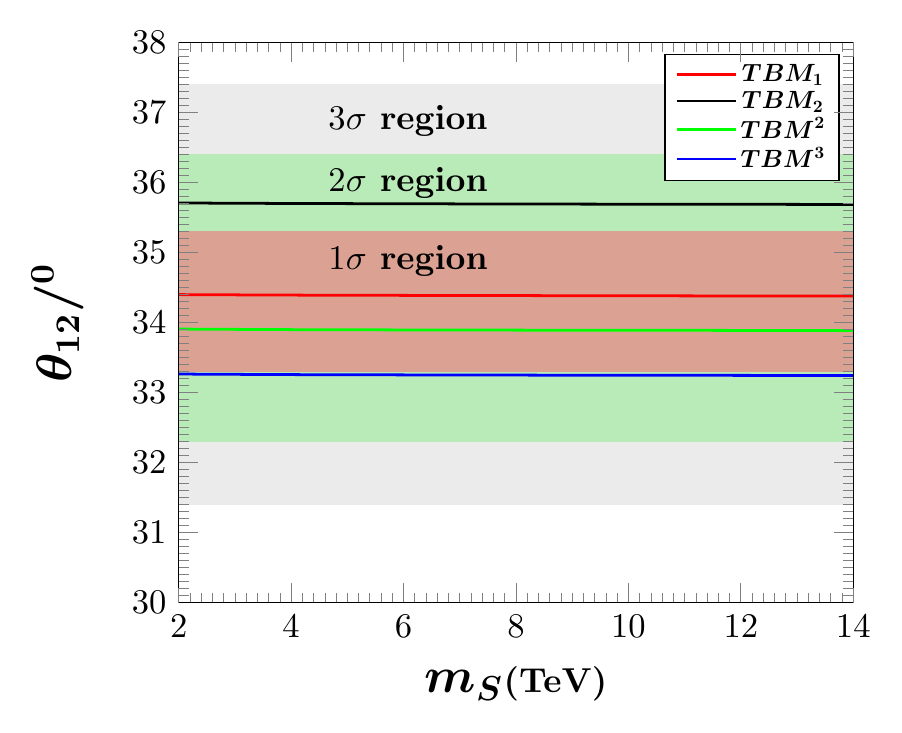}
\end{subfigure}
\begin{subfigure}{.54\textwidth}
    \includegraphics[scale=0.78]{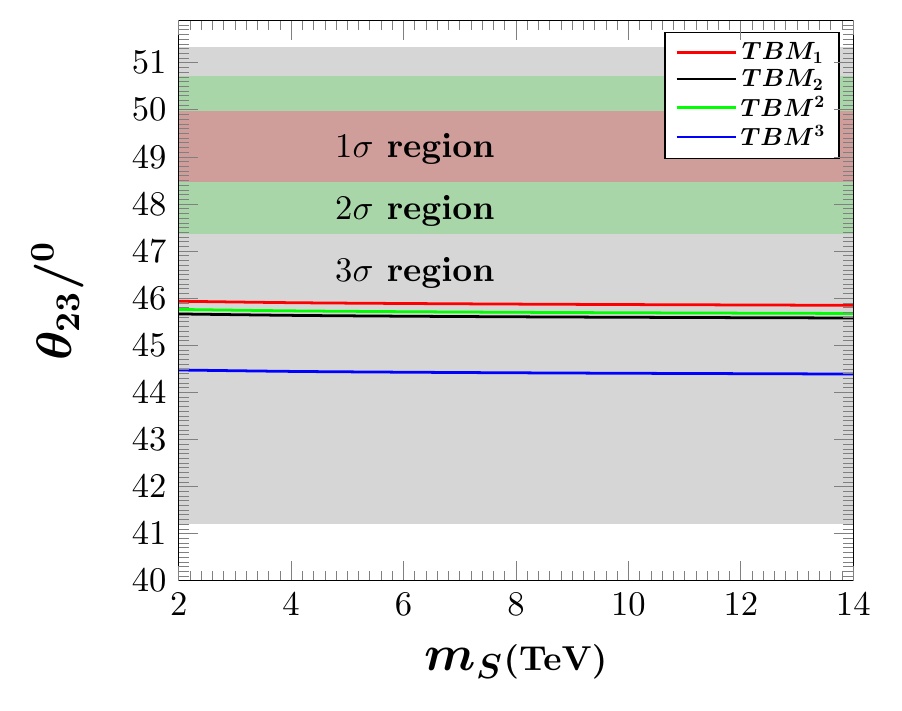}
\end{subfigure}
\begin{subfigure}{0.43\textwidth}
   \includegraphics[scale=0.78]{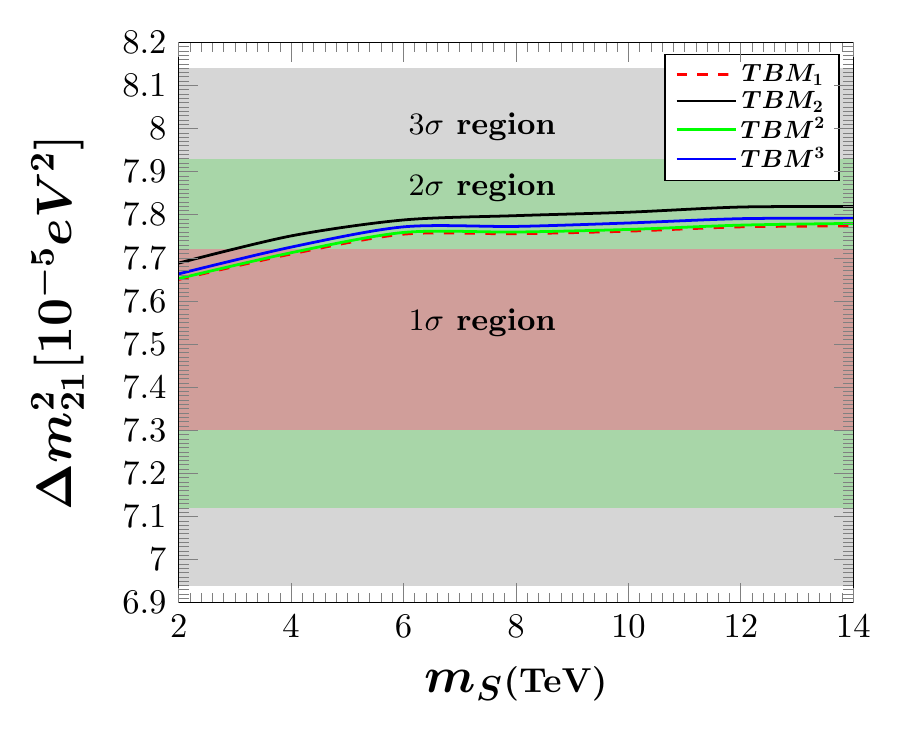}
\end{subfigure}
\begin{subfigure}{0.54\textwidth}
  \includegraphics[scale=0.78]{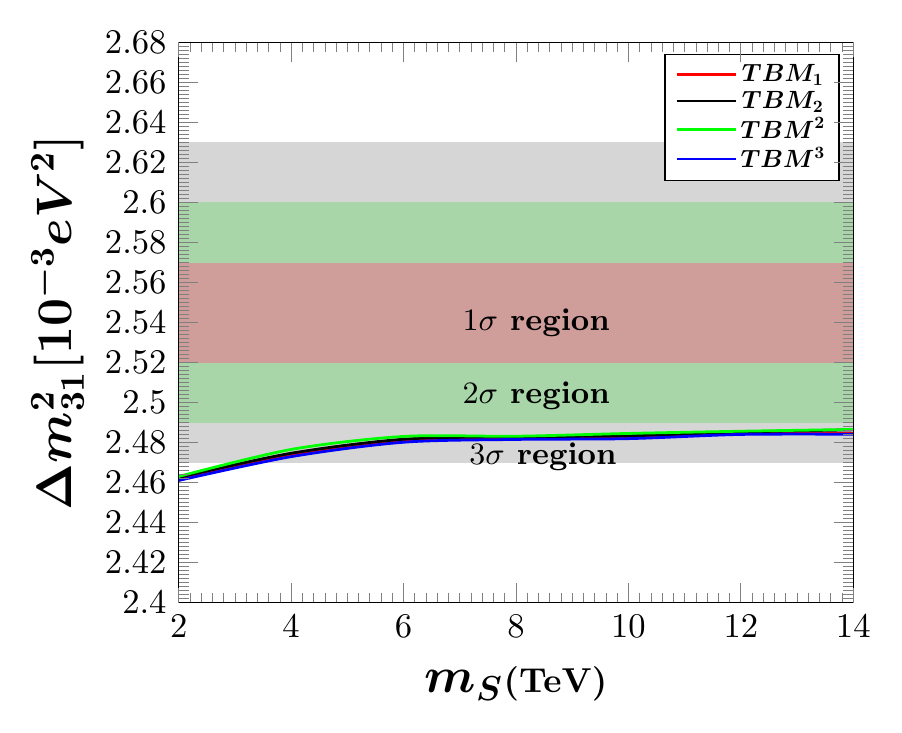}
\end{subfigure}
\begin{subfigure}{0.43\textwidth}
  \includegraphics[scale=0.78]{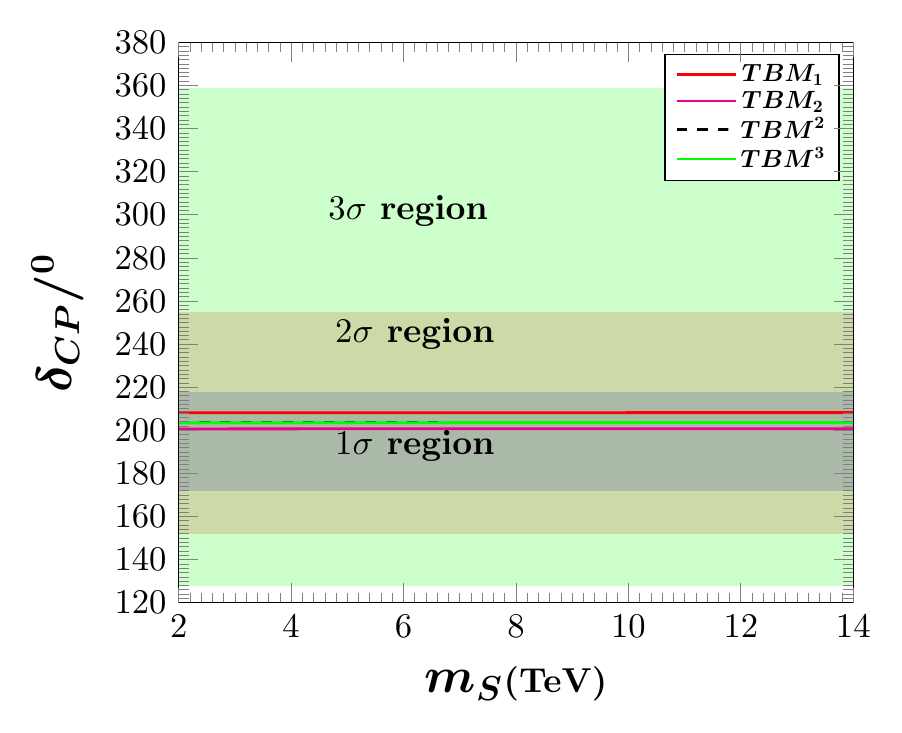}
\end{subfigure}
\caption{\footnotesize{The variation of $m_S$ for different NTBM patterns at $M_R = 10^{15}$ GeV and $\tan \beta = 30$ for NO leads to effects on the low-energy output results of $\theta_{ij}$, $|\Delta m^{2}_{ij}|$, and $\delta_{CP}$}}
\label{fig:sfig1}
\end{figure}

\begin{figure}[H]
\centering
\begin{subfigure}{0.54\textwidth}
    \includegraphics[scale=0.78]{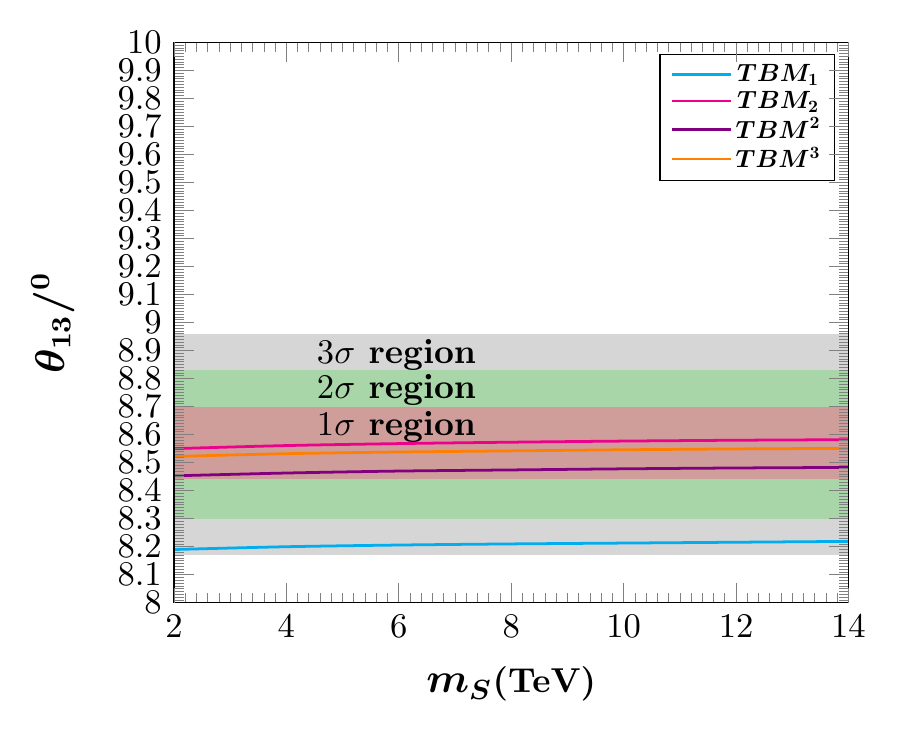}
\end{subfigure}
\begin{subfigure}{0.43\textwidth}
    \includegraphics[scale=0.78]{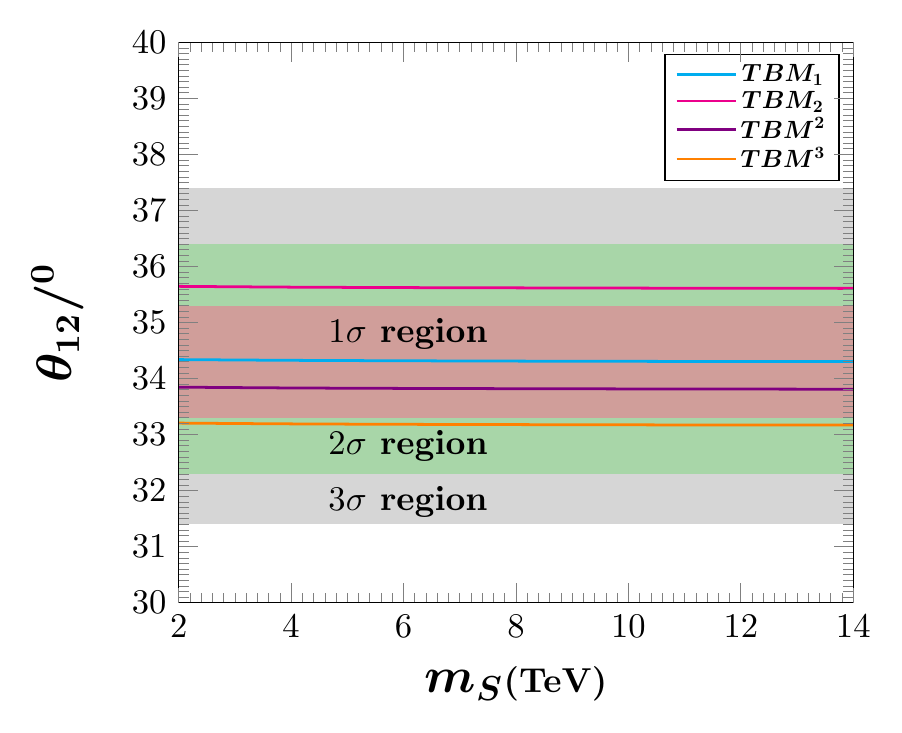}
\end{subfigure}
\begin{subfigure}{.54\textwidth}
    \includegraphics[scale=0.78]{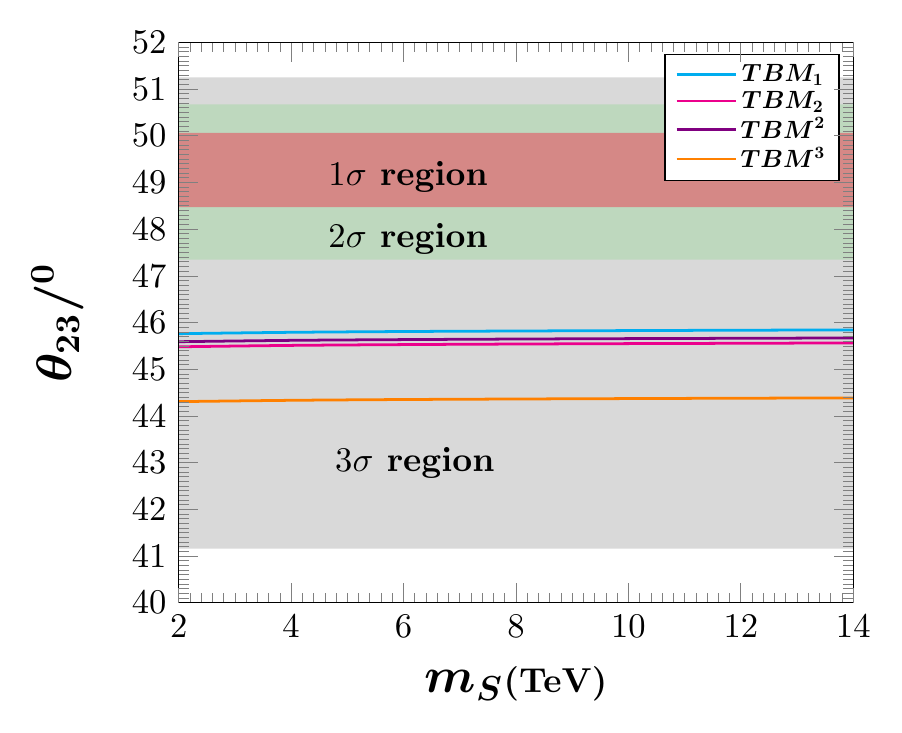}
\end{subfigure}
\begin{subfigure}{0.43\textwidth}
   \includegraphics[scale=0.78]{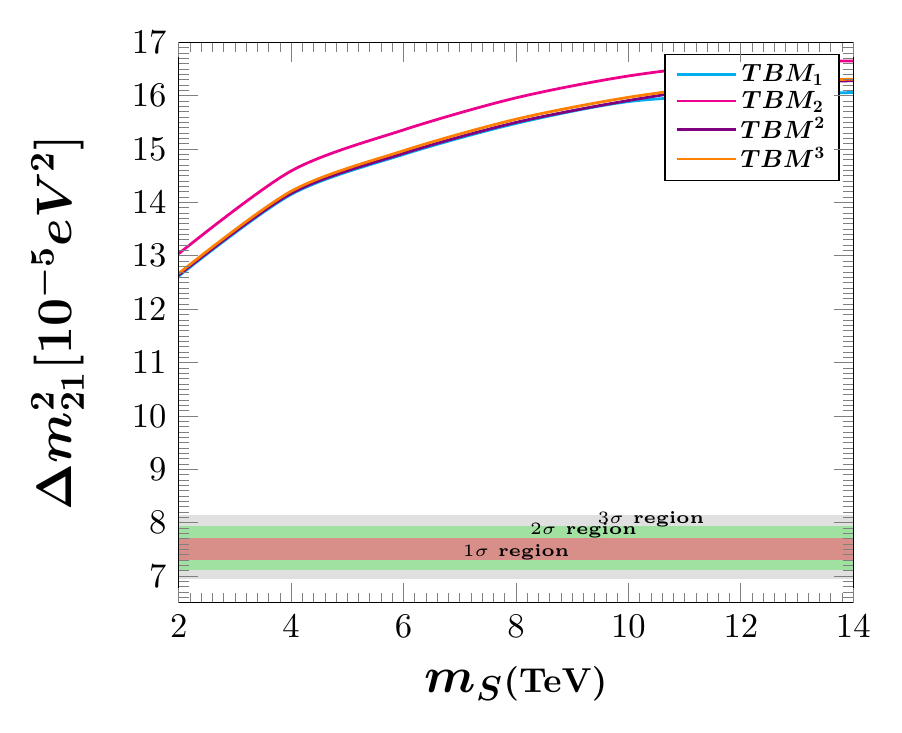}
\end{subfigure}
\begin{subfigure}{0.54\textwidth}
  \includegraphics[scale=0.78]{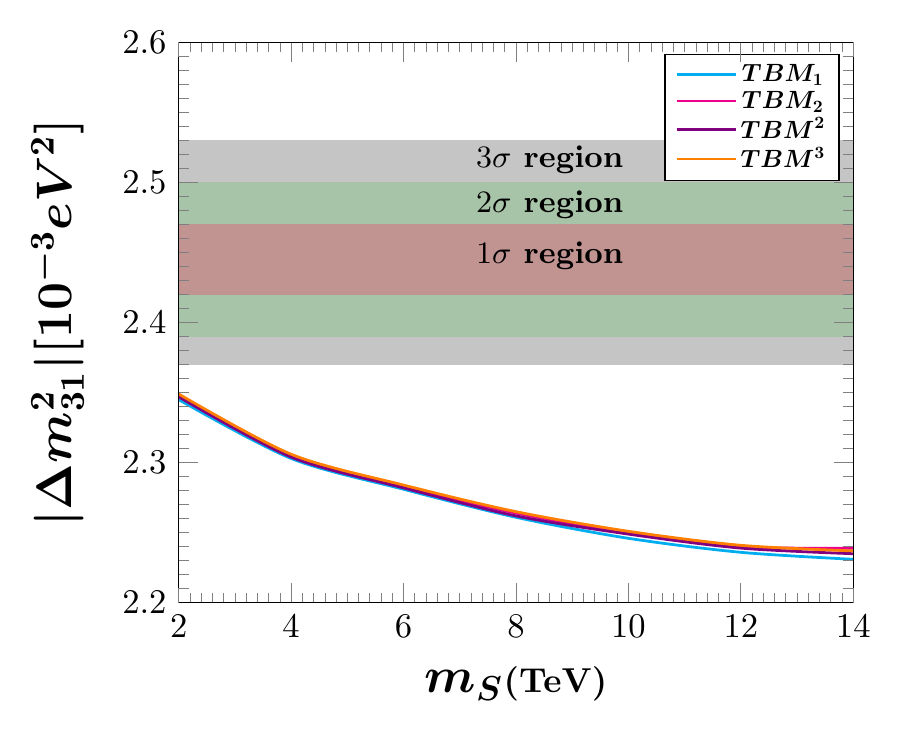}
\end{subfigure}
\begin{subfigure}{0.43\textwidth}
  \includegraphics[scale=0.78]{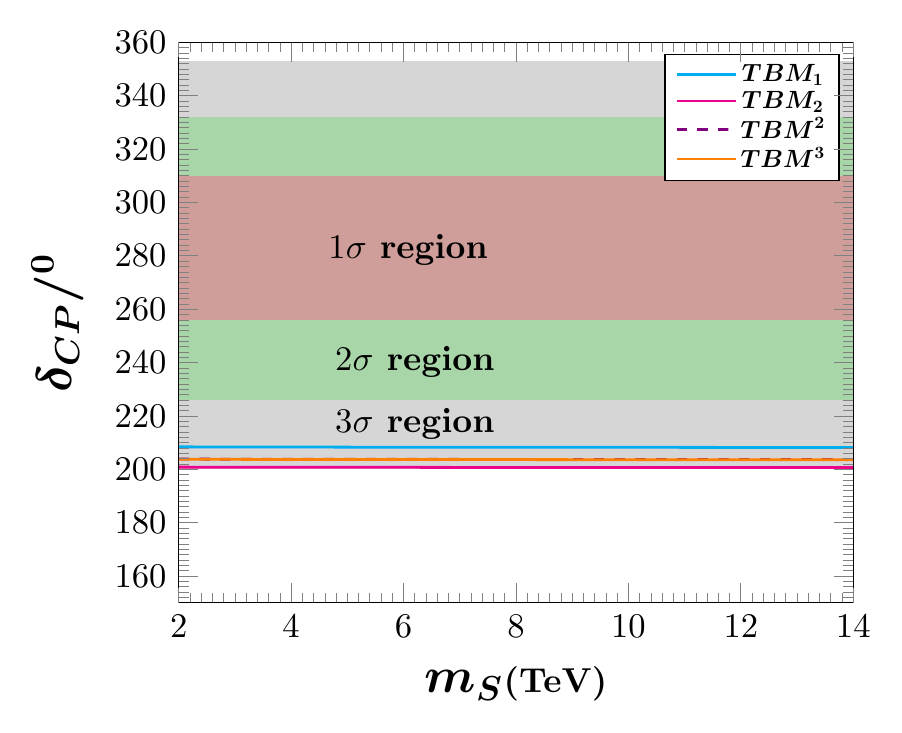}
\end{subfigure}
\caption{\footnotesize{The variation of $m_S$ for different NTBM patterns at $M_R = 10^{15}$ GeV and $\tan \beta = 50$ for IO leads to effects on the low-energy output results of $\theta_{ij}$, $|\Delta m^{2}_{ij}|$, and $\delta_{CP}$.}}
\label{fig:sfig4}
\end{figure}

   Here, we analyze the impact of varying $m_S$ while keeping the values of $M_R = 10^{15}$ GeV and $\tan \beta$ fixed at 30 and 50, respectively. We consider the effects on neutrino oscillation parameters for both NO and IO, presenting numerical data in Tables \ref{tab:4} to \ref{tab:5a} and graphical representations in Figures \ref{fig:sfig1} and \ref{fig:sfig4}.
    
   At $\tan \beta$ = 30, for NO, using the input set which is given in Table \ref{tab:i3}, it is found that all the neutrino oscillation parameters are in favor with the
latest data which are within $3\sigma$ range. All the three mixing angles decrease with increasing $m_S$  but $\Delta m_{ij}^2$ increases with increasing $m_S$. $\delta_{CP}$ maintains almost stability against the variation of $m_S$.

\begin{figure}[H]
\centering
\begin{subfigure}{0.54\textwidth}
    \includegraphics[scale=0.78]{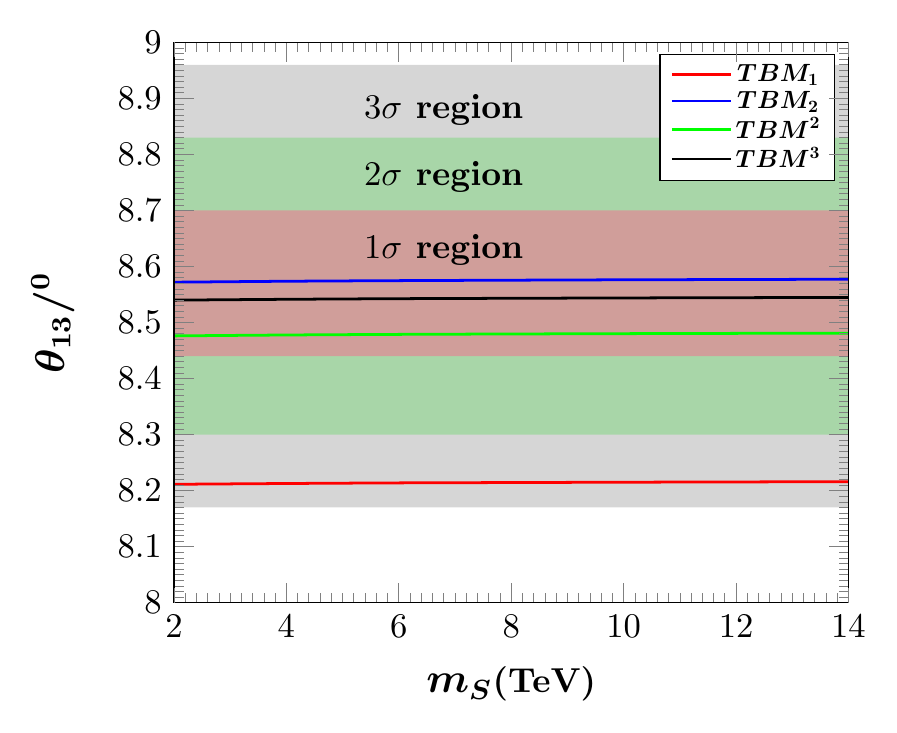}
\end{subfigure}
\begin{subfigure}{0.43\textwidth}
    \includegraphics[scale=0.78]{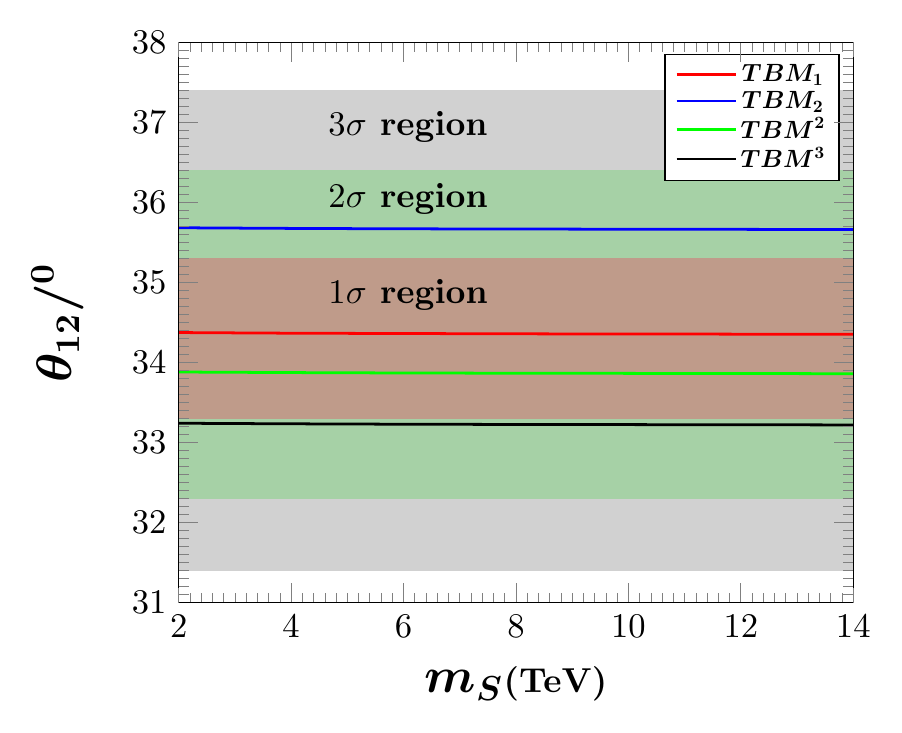}
\end{subfigure}
\begin{subfigure}{.54\textwidth}
    \includegraphics[scale=0.78]{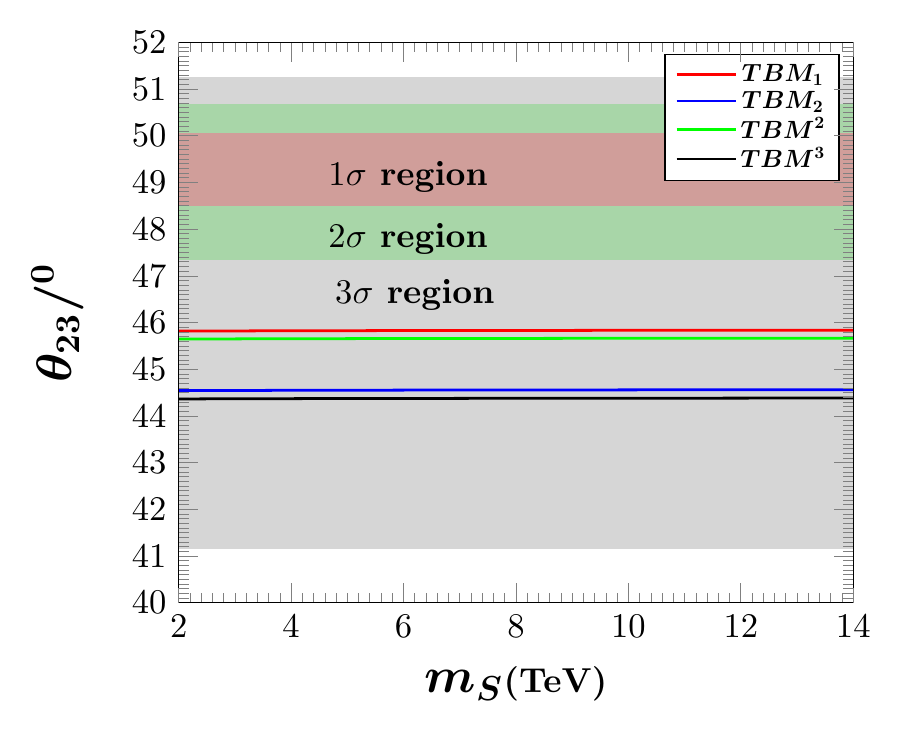}
\end{subfigure}
\begin{subfigure}{0.43\textwidth}
   \includegraphics[scale=0.78]{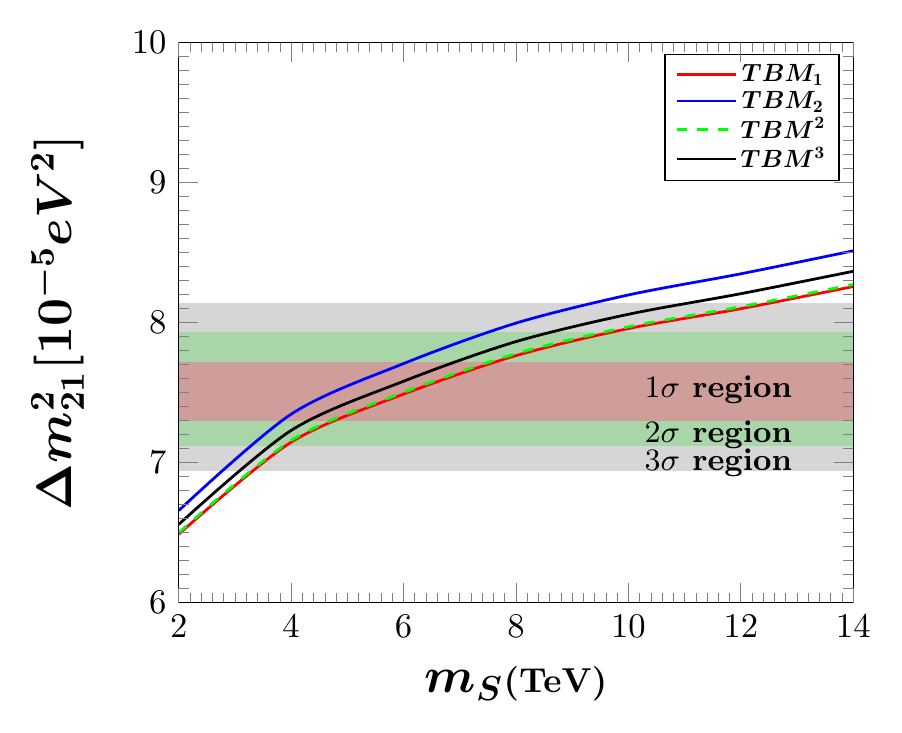}
\end{subfigure}
\begin{subfigure}{0.54\textwidth}
  \includegraphics[scale=0.78]{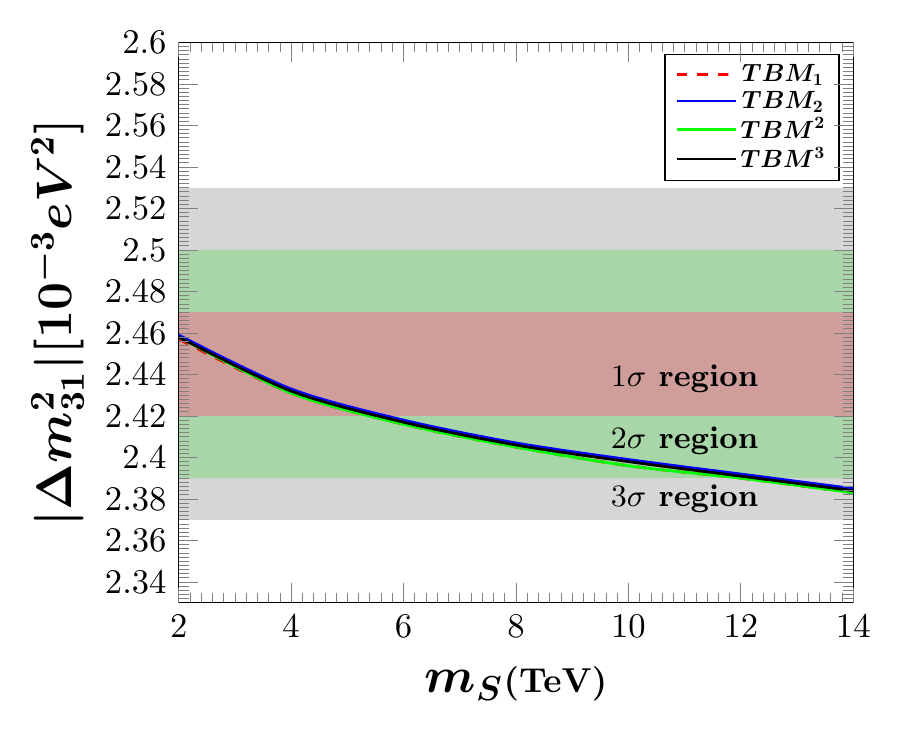}
\end{subfigure}
\begin{subfigure}{0.43\textwidth}
  \includegraphics[scale=0.78]{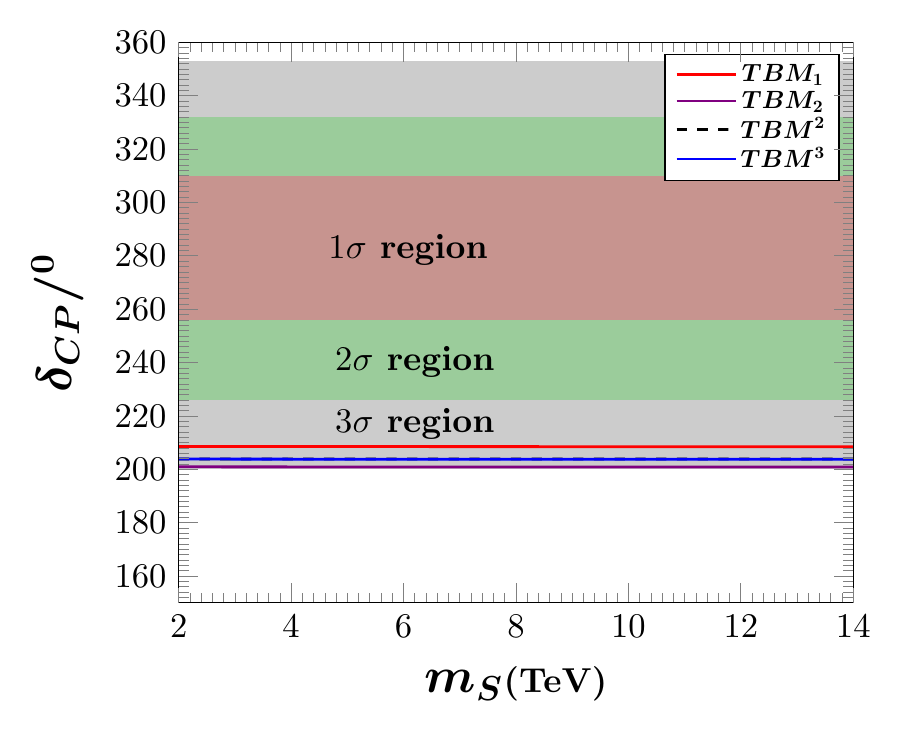}
\end{subfigure}
\caption{\footnotesize{The variation of $m_S$ for different NTBM patterns at $M_R = 10^{15}$ GeV and $\tan \beta = 30$ for IO leads to effects on the low-energy output results of $\theta_{ij}$, $|\Delta m^{2}_{ij}|$, and $\delta_{CP}$}}
\label{fig:sfig3}
\end{figure}

 At $\tan \beta$ = 30, for IO, using the input set which is given in Table \ref{tab:i3}, it is found that all the neutrino oscillation parameters are in favor with the
latest data which are within $3\sigma$ range. All the three mixing angles and $\delta_{CP}$ maintain more stability as compared to $\Delta m_{ij}^2$.  Both $\theta_{23}$ and $\theta_{13}$ increase but $\theta_{12}$ decreases against the variation of $m_S$. $\Delta m_{21}^2$  increases whereas $\Delta m_{31}^2$ decreases with increasing $m_S$ and $\Delta m_{31}^2$  maintains more stability as compared to $\Delta m_{21}^2$ at higher $m_S$. It is observed from Tables \ref{tab:4} and \ref{tab:5} and their graphical representations in Figures \ref{fig:sfig1} and \ref{fig:sfig3} that NO maintains more stability than the IO.

\begin{figure}[H]
\centering
\begin{subfigure}{0.54\textwidth}
    \includegraphics[scale=0.78]{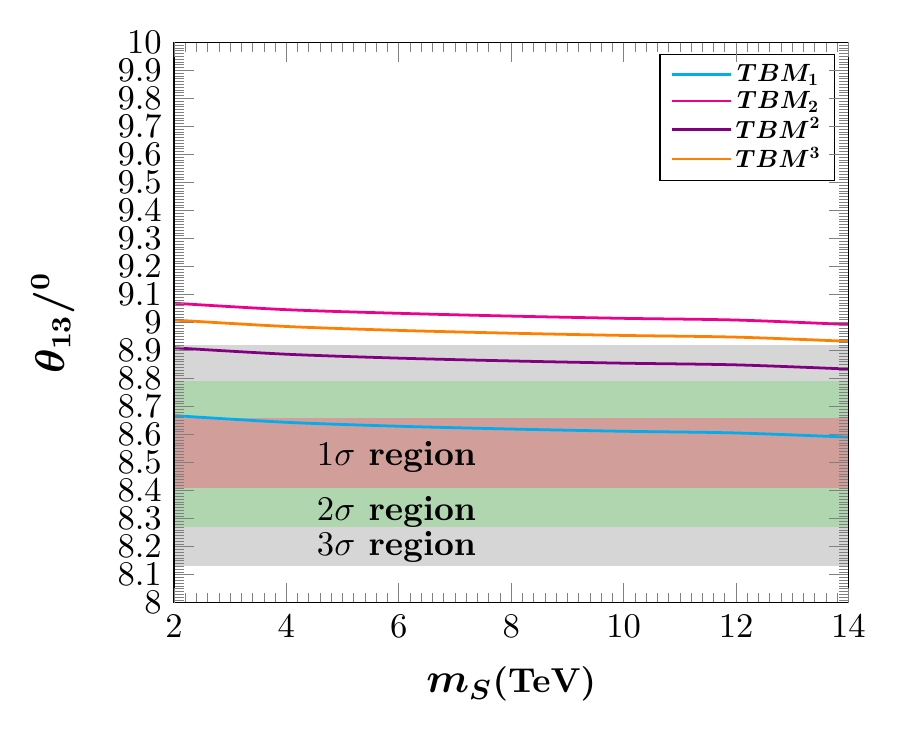}
\end{subfigure}
\begin{subfigure}{0.43\textwidth}
    \includegraphics[scale=0.78]{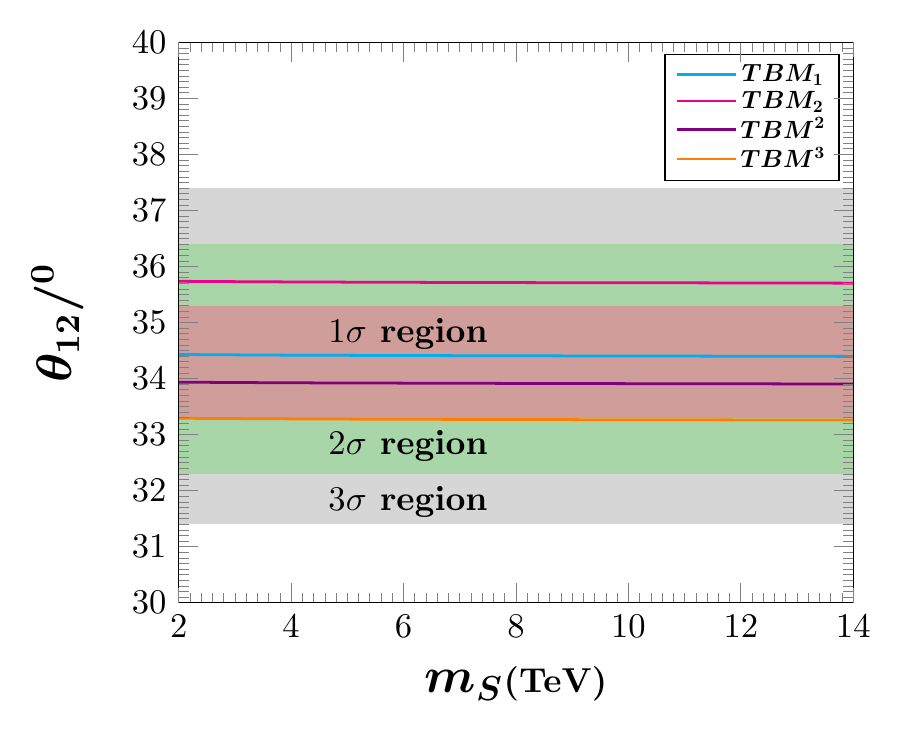}
\end{subfigure}
\begin{subfigure}{.54\textwidth}
    \includegraphics[scale=0.78]{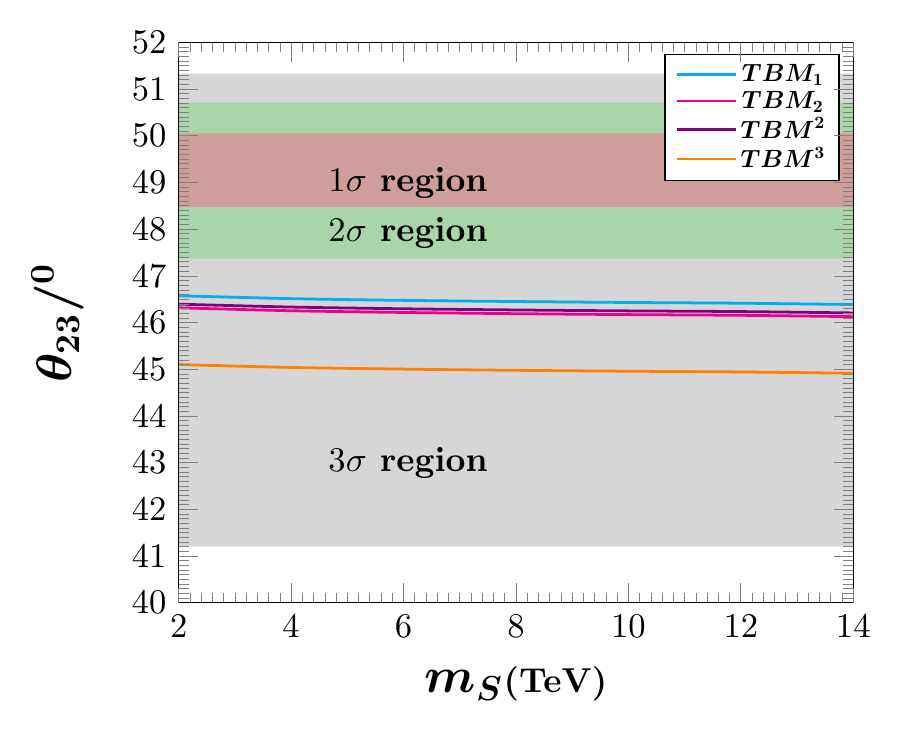}
\end{subfigure}
\begin{subfigure}{0.43\textwidth}
   \includegraphics[scale=0.78]{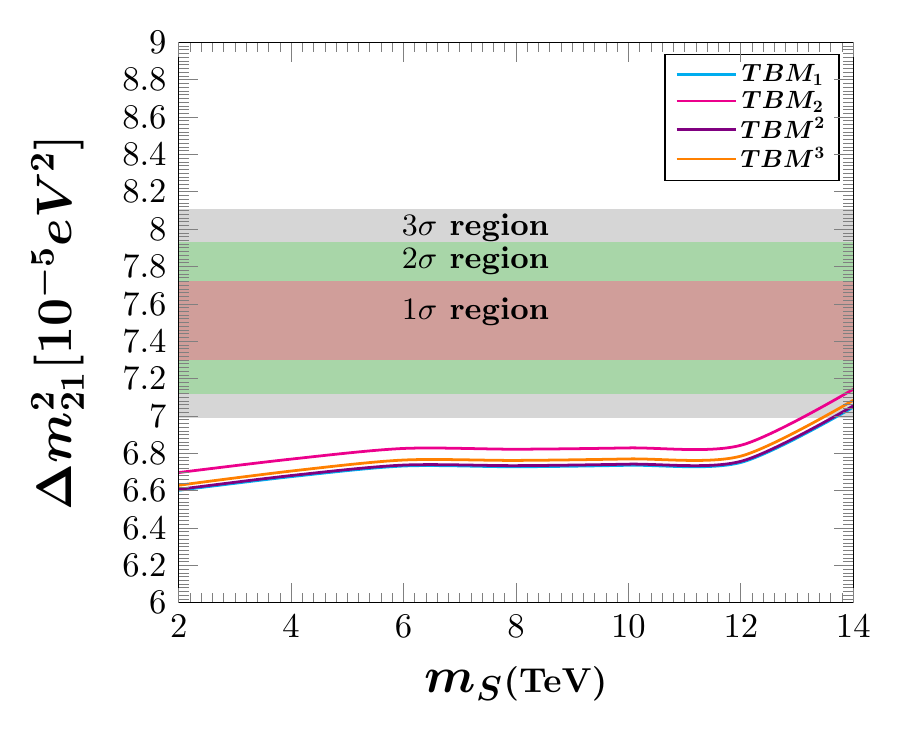}
\end{subfigure}
\begin{subfigure}{0.54\textwidth}
  \includegraphics[scale=0.78]{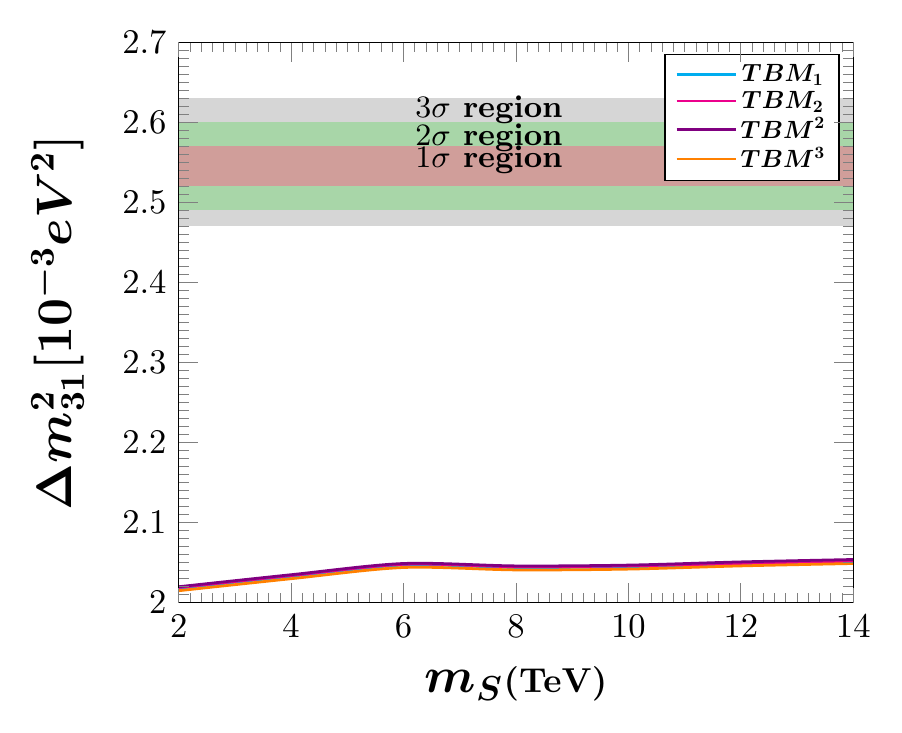}
\end{subfigure}
\begin{subfigure}{0.43\textwidth}
  \includegraphics[scale=0.78]{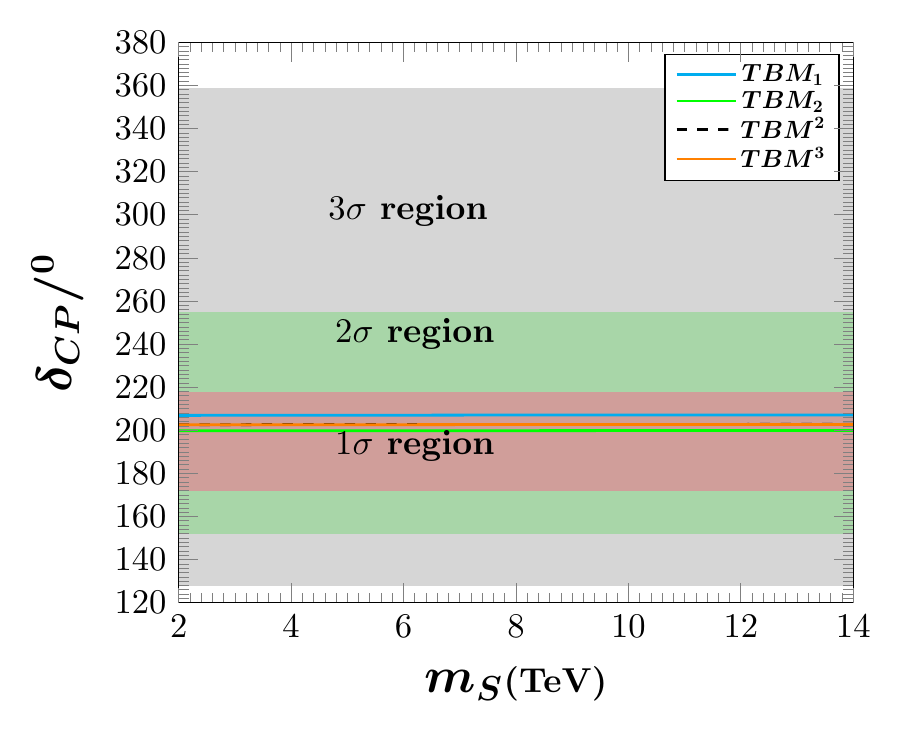}
\end{subfigure}
\caption{\footnotesize{The variation of $m_S$ for different NTBM patterns at $M_R = 10^{15}$ GeV and $\tan \beta = 50$ for NO leads to effects on the low-energy output results of $\theta_{ij}$, $|\Delta m^{2}_{ij}|$, and $\delta_{CP}$}}
\label{fig:sfig2}
\end{figure}

At $\tan \beta = 50$ and $M_R = 10^{15}$ GeV, the low energy values of the three mixing angles and $\delta_{CP}$ remain stable with the variaion of $m_S$, for both the NO and IO scenarios. However, the low energy values of the $\Delta m^2_{21}$ with the variation of $m_S$, are outside the range provided by the global fit data for IO but for NO, both the low energy values of the two-mass squared differences are within the global fit data, indicating slight preference of NO to IO. These results are presented in Tables \ref{tab:4a} and \ref{tab:5a} and are graphically represented in Figures \ref{fig:sfig2} and \ref{fig:sfig4}.
\begin{table}[H]
\centering
\begin{adjustbox}{width=1.05\textwidth}
\begin{tabular}{c|cccc|cccc|cccc}
\hline{$m_S$}& \multicolumn{4}{c|}{$\theta_{23}/^0$} & \multicolumn{4}{c|}{$\theta_{12}/^0$} & \multicolumn{4}{c}{$\theta_{13}/^0$}\\
(TeV) &$TBM_1$ &$TBM_2$ &$TBM^2$&$TBM^3$&$TBM_1$ &$TBM_2$ &$TBM^2$&$TBM^3$ &$TBM_1$ &$TBM_2$ &$TBM^2$&$TBM^3$\\
\hline
2 & 45.936&  45.666 & 45.762 & 44.476  &  34.399 & 35.707& 33.907  &33.265 & 8.269& 8.642 & 8.539 &8.601\\
4 & 45.905&  45.636 & 45.732 & 44.446 & 34.392 & 35.700 &33.899  &33.257  &8.251& 8.625 & 8.521 &8.583\\
6 & 45.888 & 45.618 & 45.714  &44.429 &  34.388 & 35.696 &33.895  &33.253 &8.241 &8.615 & 8.511 &8.573\\
8 & 45.875 & 45.605 & 45.701 & 44.416 & 34.385 &35.692 &33.892  &33.250 &8.233 &8.606 & 8.503 &8.564\\
10 & 45.865 & 45.595 & 45.691 & 44.406 & 34.382& 35.690 &33.890  &33.247 &8.226 &8.601 & 8.497 &8.558\\
12 & 45.857 & 45.587 & 45.683 & 44.398 & 34.380& 35.688 &33.888  &33.246  &8.222 &8.596 & 8.492 &8.554\\
14 & 45.849 & 45.580 & 45.676 & 44.390 &   34.379 &35.686 &33.886  &33.244 &8.217 &8.592 & 8.487 &8.549\\
\hline  
$m_S$ &  \multicolumn{4}{c|}{$\Delta m^2_{31}$ ($10^{-3} eV^2$)} & \multicolumn{4}{c|}{$\Delta m^2_{21}$ ($10^{-5} eV^2$)}& \multicolumn{4}{c}{$|\Sigma m_i|$}\\
(TeV) &$TBM_1$ &$TBM_2$ &$TBM^2$&$TBM^3$&$TBM_1$ &$TBM_2$ &$TBM^2$&$TBM^3$&$TBM_1$ &$TBM_2$ &$TBM^2$&$TBM^3$\\
\hline
 2&  2.4622 & 2.4626  & 2.4631 &2.4611 & 7.650  &7.688 & 7.652 & 7.663  & 0.0758 & 0.0758  &0.0758 & 0.0758\\
 4&  2.4736 & 2.4747  & 2.4766 &2.4730  & 7.709  &7.751 & 7.712 & 7.725&0.0755 & 0.0756  &0.0756 & 0.0756\\
 6 & 2.4807  &2.4817   &2.4831 &2.4802 & 7.755  &7.788 & 7.759 & 7.772  & 0.0754  & 0.0755  &0.0755 & 0.0755\\
 8  &2.4820  &2.4829   &2.4832 &2.4816 & 7.756  &7.798  &7.760  & 7.773 &0.0753 & 0.0754 &0.0754 &0.0754\\
 10 &2.4824  &2.4834  & 2.4846 &2.4814   & 7.762  &7.806  &7.766  & 7.781& 0.0752& 0.0753 &0.0753  &0.0753\\
 12 &2.4845  &2.4853  & 2.4857 &2.4820   & 7.772  &7.818  &7.776 & 7.791& 0.0751  & 0.0752  &0.0752 &0.0752\\
 14 &2.4846  &2.4864  & 2.4866 &2.4842  &7.774  &7.819  &7.780  & 7.792 &0.0750  &0.0751  &0.0751  &0.0751\\
\hline 
\end{tabular} 
\end{adjustbox}
\caption{ \footnotesize{Effects on the output of $\theta_{ij}$, $\Delta m^2_{ij}$ and $|\Sigma m_i|$ at low energy scale, on varying $m_S$ for four different NTBM patterns (NO) at $\tan\beta =30$ and $M_R = 10^{15}$ GeV.}}
\label{tab:4}
\end{table}

\begin{table}[H]
\centering
\begin{adjustbox}{width=1.05\textwidth}
\begin{tabular}{c|cccc|cccc|cccc}
\hline{$m_S$}& \multicolumn{4}{c|}{$\theta_{23}/^0$} & \multicolumn{4}{c|}{$\theta_{12}/^0$} & \multicolumn{4}{c}{$\theta_{13}/^0$}\\
(TeV) &$TBM_1$ &$TBM_2$ &$TBM^2$&$TBM^3$&$TBM_1$ &$TBM_2$ &$TBM^2$&$TBM^3$ &$TBM_1$ &$TBM_2$ &$TBM^2$&$TBM^3$\\
\hline
2  &  46.574   &  46.316  &  46.394  &  45.101  &34.430  &35.736  &33.936 & 33.292  & 8.668  & 9.069   &8.910  & 9.009 \\
4  &    46.512  &   46.253  &  46.332  &  45.039   & 34.418  & 35.725  & 33.925  & 33.281  & 8.644  & 9.046  & 8.887  & 8.986 \\
6  &   46.475   &  46.216  &  46.295   & 45.003  & 34.412 &  35.718  & 33.919  & 33.274   & 8.630   & 9.033  &  8.873   & 8.972\\
8  &   46.450   &  46.190  &  46.270   & 44.977  & 34.407  & 35.713  & 33.914  & 33.270 &  8.620 &  9.023   & 8.863  & 8.962\\
10 &   46.430   &  46.170  &  46.250  &  44.958 & 34.403 & 35.710 & 33.910  & 33.266  & 8.612   & 9.015   & 8.855   & 8.954\\
12 &   46.414   &  46.155  &  46.235  &  44.942 & 34.400 & 35.707 & 33.907  & 33.263  & 8.606   & 9.009   & 8.849  &  8.948\\
14  &  46.386   &  46.126   & 46.206  &  44.914 & 34.396 & 35.703 & 33.903  & 33.259  &  8.591  & 8.993  & 8.834  &  8.933\\
\hline 
$m_S$ &  \multicolumn{4}{c|}{$\Delta m^2_{31}$ ($10^{-3} eV^2$)} & \multicolumn{4}{c|}{$\Delta m^2_{21}$ ($10^{-5} eV^2$)}& \multicolumn{4}{c}{$|\Sigma m_i|$}\\
(TeV) &$TBM_1$ &$TBM_2$ &$TBM^2$&$TBM^3$&$TBM_1$ &$TBM_2$ &$TBM^2$&$TBM^3$&$TBM_1$ &$TBM_2$ &$TBM^2$&$TBM^3$\\
\hline

2 & 2.018 & 2.019 & 2.020 & 2.015 &  6.602 &  6.697 &  6.607 &  6.629 & 0.0688 & 0.0685 & 0.0686 & 0.0688 \\
4 &  2.033 & 2.034 & 2.035 &  2.030 & 6.675 & 6.769 & 6.681 & 6.705 & 0.0687 & 0.0687 & 0.0687 & 0.0686\\
6 &  2.047 & 2.048 & 2.049 &  2.044 &  6.732 & 6.826 & 6.737 & 6.764 & 0.0687 & 0.0687 & 0.0687 & 0.0686\\
8  & 2.044 & 2.045 & 2.046 & 2.041  & 6.728  & 6.822 & 6.734 &  6.762 & 0.0684 & 0.0684 & 0.0684 & 0.0683\\
10 & 2.045 & 2.046 & 2.047 & 2.042 &  6.736 & 6.829  & 6.742 & 6.770 & 0.0683 & 0.0683 & 0.0683 & 0.0683\\
12 & 2.048 & 2.049 & 2.051 & 2.046 &  6.751 & 6.843  & 6.757 &  6.785 & 0.0683 & 0.0683 & 0.0683 & 0.0683\\
14 & 2.051 & 2.052 & 2.054 & 2.049 & 7.047  & 7.143 & 7.054 & 7.084 & 0.0682 & 0.0682 & 0.0682 & 0.0682\\
\hline
\end{tabular} 
\end{adjustbox}
\caption{\footnotesize{Effects on the output of $\theta_{ij}$, $\Delta m^2_{ij}$ and $|\Sigma m_i|$ at low energy scale, on varying $m_S$ for four different NTBM patterns (NO) at $\tan\beta =50$ and $M_R = 10^{15}$ GeV.}}
\label{tab:4a}
\end{table}

\begin{table}[H]
\centering
\begin{adjustbox}{width=1.05\textwidth}
\begin{tabular}{c|cccc|cccc|cccc}
\hline{$m_S$}& \multicolumn{4}{c|}{$\theta_{23}/^0$} & \multicolumn{4}{c|}{$\theta_{12}/^0$} & \multicolumn{4}{c}{$\theta_{13}/^0$}\\
(TeV) &$TBM_1$ &$TBM_2$ &$TBM^2$&$TBM^3$&$TBM_1$ &$TBM_2$ &$TBM^2$&$TBM^3$ &$TBM_1$ &$TBM_2$ &$TBM^2$&$TBM^3$\\
\hline
2  & 45.8174   &  44.5423 &  45.6469 & 44.3643 & 34.375 & 35.684 & 33.882 &  33.242 &  8.2114 & 8.5720 & 8.4762 & 8.5400\\
4  & 45.8240 & 44.5490  & 45.6535 & 44.3708  & 34.367  & 35.676 & 33.875  & 33.235  & 8.2128 & 8.5737 & 8.4778  & 8.5415\\
6  & 45.8283 & 44.5533  & 45.6577 & 44.3749  & 34.363  & 35.671 & 33.870  & 33.230  & 8.2139 & 8.5748 &  8.4789  & 8.5426\\
8  & 45.8311 & 44.5563  & 45.6605 & 44.3778  & 34.360  & 35.668 & 33.867  & 33.227   & 8.2145 & 8.5756 & 8.4796  & 8.5433\\
10 & 45.8334 & 44.5585  & 45.6627 & 44.3802  & 34.357  & 35.666 & 33.865  & 33.225 & 8.2151 & 8.5763  & 8.4802  &8.5439 \\
12 & 45.8355  & 44.5607  & 45.6648 & 44.3819 & 34.355  & 35.664 & 33.863  & 33.223 & 8.2156 & 8.5768  & 8.4807 & 8.5444\\
14 & 45.8371  & 44.5624  & 45.6664 & 44.3835  & 34.353  & 35.662 & 33.861 & 33.221 & 8.2160 &8.5773  & 8.4811 & 8.5448\\
\hline 
$m_S$ &  \multicolumn{4}{c|}{$|\Delta m^2_{31}|$ ($10^{-3} eV^2$)} & \multicolumn{4}{c|}{$\Delta m^2_{21}$ ($10^{-5} eV^2$)}& \multicolumn{4}{c}{$|\Sigma m_i|$}\\
(TeV) &$TBM_1$ &$TBM_2$ &$TBM^2$&$TBM^3$&$TBM_1$ &$TBM_2$ &$TBM^2$&$TBM^3$&$TBM_1$ &$TBM_2$ &$TBM^2$&$TBM^3$\\
\hline
 2  & 2.457 & 2.459 & 2.458 & 2.458  & 6.489 & 6.658 & 6.499 & 6.559 &0.1096&0.1096&0.1096&0.1096\\
4  & 2.431 & 2.433 & 2.431 & 2.432 & 7.146 & 7.346  & 7.158 & 7.230&0.1091&0.1091&0.1091&0.1091\\
6  & 2.416 &  2.418 & 2.416 & 2.417 &  7.489 &  7.707 & 7.501 &  7.581&0.1089&0.1089&0.1089&0.1089\\
8 & 2.405 & 2.407 & 2.405 & 2.406  & 7.764  & 7.996  & 7.778  & 7.864 &0.1086&0.1087&0.1086&0.1086\\
10 & 2.396 & 2.399 & 2.396 & 2.398 & 7.957  & 8.197  & 7.970  & 8.059&0.1085 &0.1085 &0.1085&0.1085\\
12 & 2.390 & 2.392 & 2.390 & 2.391 & 8.099  & 8.348  & 8.113  & 8.206&0.1084&0.1084&0.1084&0.1084\\
14 & 2.383 & 2.385 & 2.383 & 2.384  & 8.257  & 8.512  & 8.272  & 8.366&0.1082&0.1083&0.1082&0.1083\\
\hline 
\end{tabular} 
\end{adjustbox}
\caption{\footnotesize{Effects on the output of $\theta_{ij}$, $|\Delta m^2_{ij}|$ and $|\Sigma m_i|$ at low energy scale, on varying $m_S$ for four different NTBM patterns (IO) at $\tan\beta =30$ and $M_R = 10^{15}$ GeV.}}
\label{tab:5}
\end{table}

\begin{table}[H]
\centering
\begin{adjustbox}{width=1.05\textwidth}
\begin{tabular}{c|cccc|cccc|cccc}
\hline{$m_S$}& \multicolumn{4}{c|}{$\theta_{23}/^0$} & \multicolumn{4}{c|}{$\theta_{12}/^0$} & \multicolumn{4}{c}{$\theta_{13}/^0$}\\
(TeV) &$TBM_1$ &$TBM_2$ &$TBM^2$&$TBM^3$&$TBM_1$ &$TBM_2$ &$TBM^2$&$TBM^3$ &$TBM_1$ &$TBM_2$ &$TBM^2$&$TBM^3$\\
\hline
2  & 45.762  & 45.481   & 45.590  &  44.307  &   34.339  & 35.646  & 33.847 &  33.206  & 8.190 &   8.550  &  8.453  &  8.522 \\
4  &  45.791  & 45.511  &  45.619 &   44.335 &   34.326 &  35.632  & 33.833 &  33.192  &  8.200 &   8.561  &  8.463   & 8.532 \\
6  &  45.808  & 45.528 &   45.635  &  44.352  & 34.318  & 35.624  & 33.825 &  33.184  &  8.206  &  8.568  &  8.470  &  8.538 \\
8  &  45.819  & 45.540  &  45.647  &  44.363  &  34.312 &  35.619 &  33.820  & 33.179 &  8.210  &  8.573  &  8.474  &  8.542 \\
10  & 45.828  & 45.549  &  45.656  &  44.372  &  34.308 &  35.615 &  33.816  & 33.175  &  8.213  &  8.577  &  8.478  &  8.546 \\
12 &  45.836 &  45.556  &  45.663  &  44.380  &  34.305  & 35.611  & 33.813 &  33.171  &  8.216  &  8.580  &  8.481  &  8.549 \\
14  & 45.842  & 45.564  &  45.670  &  44.386  &  34.303 &  35.610  & 33.811 &  33.170  &  8.219  &  8.582  &  8.483  &  8.551 \\
\hline
$m_S$ &  \multicolumn{4}{c|}{$|\Delta m^2_{31}|$ ($10^{-3} eV^2$)} & \multicolumn{4}{c|}{$\Delta m^2_{21}$ ($10^{-5} eV^2$)}& \multicolumn{4}{c}{$|\Sigma m_i|$}\\
(TeV) &$TBM_1$ &$TBM_2$ &$TBM^2$&$TBM^3$&$TBM_1$ &$TBM_2$ &$TBM^2$&$TBM^3$&$TBM_1$ &$TBM_2$ &$TBM^2$&$TBM^3$\\
\hline
2 &   2.345 &  2.348 &  2.347 & 2.349&  12.62 &  13.04 &  12.64 & 12.67 & 0.1084 & 0.1085 & 0.1084 & 0.1084\\
4 &   2.303 &  2.305 & 2.304 & 2.306 &   14.15 &   14.59 &  14.17 & 14.21 & 0.1078 & 0.1079 & 0.1079 & 0.1079 \\
6 &  2.281 & 2.283 & 2.282 & 2.284  &  14.90 &  15.36 &  14.92 & 14.97& 0.1076 & 0.1077 & 0.1076 & 0.1076\\
8 &  2.261 & 2.264 & 2.262 & 2.265  & 15.48  & 15.96 &  15.50 & 15.56& 0.1074 & 0.1075 & 0.1074 & 0.1074 \\
10 &  2.246 & 2.250 &  2.249 & 2.251 & 15.89 &  16.37 &  15.91 &  15.97& 0.1072 & 0.1073 & 0.1072 & 0.1072 \\
12 &  2.236 & 2.240 & 2.239 &  2.241 &   16.01 &  16.61 &  16.21 & 16.23& 0.1071 & 0.1072 & 0.1071 & 0.1071\\
14 & 2.231 & 2.239 & 2.235 & 2.237 & 16.06  & 16.65 &  16.28 &  16.31 & 0.1070 & 0.1071 & 0.1070 & 0.1077\\
\hline 
\end{tabular} 
\end{adjustbox}
\caption{ \footnotesize{Effects on the output of $\theta_{ij}$, $|\Delta m^2_{ij}|$ and $|\Sigma m_i|$ at low energy scale, on varying $m_S$ for four different NTBM patterns (IO) at $\tan\beta =50$ and $M_R = 10^{15}$ GeV.}}
\label{tab:5a}
\end{table}


If we consider the position (left or right) of the multiplication by a unitary rotation matrix to tribimaximal (TBM) mixing, assuming these unitary rotation matrices are originated from charged lepton mass matrices, $TBM^2$  and $TBM^3$ are acceptable as compared to $TBM_1$  and $TBM_2$, as the position of the unitary matrices should be on the left side of TBM matrix. For example, $U^3 = R_{12} V_{TBM}$. Further if we also consider
the graphical analysis of $\kappa_1$ and $\kappa_2$ for obtaining the best fit in both NO and IO, $TBM^3$ seems to be better than other three types of NTBM, as the ranges of $\kappa_1$ and $\kappa_2$ in $TBM^3$, are common for both NO and IO as shown in Figures \ref{fig:figab} and \ref{fig:figac} for four different NTBM patterns. For other remaining three cases of NTBM, the ranges of $\kappa_1$ and $\kappa_2$ are different in NO and IO. Considering the above two points, our analysis shows that $TBM^3$ is the best candidate.

\section{Discussion and Conclusions}
To summarize, we impose the following conditions to obtain the best fit pattern models among the NTBM.
\begin{enumerate}[label=\roman*.]
\item The input of the sum of three neutrino masses should satisfy the latest PLANCK cosmological data $\Sigma |m_i|<0.12$ eV.
\item We apply the conditions $\sin \delta_{CP} > 0 (NO)$ and  $\sin \delta_{CP} < 0 (IO)$ respectively, in order to constraint the two free parameters  $\kappa_1$ and  $\kappa_2$ for all the NTBM scenarios.
\item  We take the values of $\kappa_1$ and  $\kappa_2$ which give the latest values of three mixing angles given by the latest global fit data.
\item  We take different values of $\kappa_1$ and  $\kappa_2$ which lie within the allowed-regions as depicted in Figures \ref{fig:figab} and \ref{fig:figac} for four different NTBM patterns for both NO and IO.
\end{enumerate}
We study the stability for four different NTBM patterns at fixed value of $M_R = 10^{15}$ GeV and two different values of $\tan \beta$ (30, 50) for both NO and IO. 

\textbf{Case A ($\tan \beta = 30$)}: For NO, we have studied the stability for four patterns of NTBM using RGEs against the variation of $m_S$. There is a mile decrease of  the mixing angles $\theta_{ij}$ and $\delta_{CP}$ with the increase of $m_S$ (2 TeV - 14 TeV). These are found to lie within $3 \sigma$ ranges of observational data.  $\Delta m_{ij}^2$ and $\delta_{CP}$ increase with increasing $m_S$. The low energy values of $\Delta m_{21}^2$ is found to lie within $2 \sigma$ whereas  $\Delta m_{31}^2$ lies within $3 \sigma$ except for low values of $m_S$.

Similarly, for IO the low energy values of  $\Delta m_{31}^2$  and $\delta_{CP}$ decrease with the increase of $m_S$ (2 TeV - 14 TeV) which  are found to  lie within $3 \sigma$ ranges. $\Delta m_{21}^2$ is found to lie within $3 \sigma$ ranges, which increases with increasing $m_S$. $\theta_{ij}$ (except $\theta_{12}$) increases slightly with the increase of $m_S$.

In short, it is found that NTBM mixing patterns maintain stability under radiative corrections with the variation of $m_S$ for  Normal ordering case at the fixed value of seesaw scale $M_R$. All the neutrino oscillation parameters receive varying radiative corrections irrespective of the $m_S$ values at the electroweak scale, which are all within $3\sigma$ range of the latest global fit data. NO maintains more stabilty as compared to IO with increasing $m_S$. All the four patterns of NTBM are found to be stable with the variation of $m_S$ under radiative corrections in MSSM for both NO $(\sin \delta > 0)$ and IO $(\sin \delta < 0)$. If we consider the graphical analysis for $\kappa_1$ and $\kappa_2$ both for NO and IO as depicted in Figures \ref{fig:figab} and \ref{fig:figac}, $TBM^3$ is the best candidate, since it the most consistent one among the four NTBM cases.

\textbf{Case B ($\tan \beta = 50$)}: For both NO and IO, the low energy values of all the three mixing angles with the variation of $m_S$, are within the $3\sigma$ range. They remain stable with the variation of $m_S$. For IO, the low energy values of the $\Delta m^2_{21}$ with the variation of $m_S$, fall outside the range of the global fit data. However for NO, both the low energy values of the two-mass squared differences, fall within $3 \sigma$ range of the global fit data. This indicates slight preference of NO to IO in our numerical analysis. Additionally, all of the low energy values of the neutrino oscillation parameters undergo distinct radiative corrections. The graphical representations can be seen in Figures \ref{fig:sfig2} and \ref{fig:sfig4}, accompanied by the numerical data presented in Tables \ref{tab:4a} and \ref{tab:5a}.

\section*{APPENDIX A}
\subsubsection*{RGEs for gauge couplings \cite{rg1}:}
The two loop RGEs for gauge couplings  are given by
\begin{equation}
 \frac{dg_i}{dt}= \frac{b_i}{16 \pi^2} g_i^3  + \frac{1}{(16 \pi^2)^2}\biggl[ \sum_{\mathclap{j=1}}^{3}b_{ij} g_i ^3 g_j ^2 - \sum_{\mathclap{j=t, b, \tau}} a_{ij}g_i ^3 h_j ^2 \biggl],
\end{equation} 

\bigskip
where t = ln $\mu$ and $ b_i ,b_{ij}, a_{ij}$ are $\beta$ function coefficients in MSSM, 
$$ b_{i} =
\left(\begin{array}{ccc}
6.6,  & 1.0,  &   -3.0
\end{array}\right) ,
b_{ij}=
\left(\begin{array}{ccc}
7.96 & 5.40 & 17.60 \\ 
1.80 & 25.00 & 24.00 \\ 
2.20 & 9.00 & 14.00
\end{array}\right),$$
$$ a_{ij}=
\left(\begin{array}{ccc}
5.2 & 2.8 & 3.6 \\ 
6.0 & 6.0 & 2.0 \\ 
4.0 & 4.0 & 0.0
\end{array}\right)$$ 
 and, for NOn-supersymmetric case, we have\\
 
$$ 
b_i=
\left(\begin{array}{ccc}
4.100,  & -3.167,  &   -7.00
\end{array}\right) ,
b_{ij}=
\left(\begin{array}{ccc}
3.98 & 2.70 & 8.8 \\ 
0.90 & 5.83 & 12.0 \\ 
1.10 & 4.50 & -26.0
\end{array}\right),$$  and 
$$a_{ij}=
\left(\begin{array}{ccc}
0.85 & 0.5 & 0.5 \\
1.50 & 1.5 & 0.5 \\ 
2.00 & 2.0 & 0.0
\end{array}\right).$$  \\

\subsubsection*{Two-loop RGEs for Yukawa couplings and quartic Higgs coupling \cite{rg1}:} 
  For MSSM,
\begin{eqnarray}
\frac{dh_t}{dt} &=&\frac{h_t}{16 \pi^2}\biggl(6h_{t}^2 + h_{b}^2 -\sum_{\mathclap{i=1}}^{3} c_i g_{i}^2 \biggl)
+\frac{h_t}{(16 \pi^2)^2}\biggl[\sum_{\mathclap{i=1}}\biggl(c_i b_i +\frac{c_{i}^2}{2}\biggl)g_{i}^4  + g_{1}^2 g_{2}^2 \biggl. \nonumber\\
& & + \frac{136}{45} g_{1}^2 g_{3}^2 + 8 g_{2}^2 g_{3}^2 +\biggl(\frac{6}{5}g_{1}^2+6 g_{2}^2+16g_{3}^2\biggl)h_{t}^2+ \frac{2}{5}g_{1}^2 h_{b}^2 -22h_{t}^4 \nonumber\\
&& \biggl.- 5 h_{b}^4-5h_{t}^2h_{b}^2 - h_{b}^2 h_{\tau}^2\biggl],
\end{eqnarray}

\begin{eqnarray}
\frac{dh_b}{dt} &=&\frac{h_b}{16 \pi^2}\biggl(6h_{b}^2 + h_{\tau}^{2}+ h_{t}^2- \sum_{\mathclap{i=1}}^{3} c_{i}^{'} g_{i}^2\biggl)+\frac{h_{b}}{(16 \pi^{2})^2}\biggl[ \sum_{\mathclap{i=1}} \biggl(c_{i}^{'} b_i +\frac{c{i}^{'2}}{2}\biggl) g_{i}^4 \biggl. \nonumber\\
&&+ g_{1}^2 g_{2}^2 +\frac{8}{9}g_{1}^2 g_{3}^2  + 8g_{2}^2 g_{3}^2+\biggl(\frac{2}{5} g_{1}^2 +6g_{2}^2+16 g_{3}^2\biggl) h_{b}^2 +\frac{4}{5} g_{1}^2 h_{t}^2 + \frac{6}{5}g_{1}^2 h_{\tau}^2   \nonumber\\
&& \biggl.-22h_{b}^4 - 3 h_{\tau}^4-5h_{t}^4 -5h_{b}^2 h_{t}^2 -3h_{b}^2 h_{\tau}^2\biggl],
\end{eqnarray}

\begin{eqnarray}
\frac{dh_\tau}{dt} &=&\frac{h_\tau}{16 \pi^2}\biggl(4h_{\tau}^2 +3 h_{b}^{2}- \sum_{\mathclap{i=1}}^{3} c_{i}^{''} g_{i}^2\biggl)+\frac{h_{\tau}}{(16 \pi^{2})^2}\biggl[ \sum_{\mathclap{i=1}} \biggl(c_{i}^{''} b_i +\frac{c{i}^{''2}}{2}\biggl)
 g_{i}^4 \biggl. \nonumber\\
&& \biggl.+\frac{9}{5} g_{1}^2 g_{2}^2 + \biggl(\frac{6}{5}g_{1}^2+6 g_{2}^2\biggl) h_{\tau}^2 +\biggl(\frac{-2}{5} g_{1}^2 +16g_{3}^2\biggl)h_{b}^2 +9 h_{b}^4 \biggl. \nonumber\\
&& \biggl.- 10 h_{\tau}^4-3 h_{b}^2 h_{t}^2  -9h_{b}^2 h_{\tau}^2\biggl],
\end{eqnarray}

where  $$c_{i}=
\left(\begin{array}{ccc}
\frac{13}{15}, & 3, & \frac{16}{13}
\end{array}\right) , 
c_{i}^{'}=
\left(\begin{array}{ccc}
\frac{7}{15}, & 3, & \frac{16}{3}  
\end{array}\right)$$  and
$$c_{i}^{''}=
\left(\begin{array}{ccc}
\frac{9}{5} ,& 3 ,& 0  
\end{array}\right).$$  
For NOn-supersymmetric case,

\begin{eqnarray}{\label{eq:t}}
\frac{dh_t}{dt} &=&\frac{h_t}{16 \pi^2}\biggl(\frac{3}{2} h_{t}^2 - \frac{3}{2} h_{b}^2 +Y_{2}(S) - \sum_{\mathclap{i=1}}^{3} c_i g_{i}^2\biggl) + \frac{h_t}{(16 \pi^2)^2}\biggl[\biggl(\frac{1187}{600}\biggl)g_{1}^4 - \frac{23}{4} g_{2}^4  \biggl. \nonumber\\
&&\biggl.- 108 g_{3}^4-\frac{9}{20} g_{1}^2 g_{2}^2+ \frac{19}{15} g_{1}^{2} g_{3}^2 +9 g_{3}^2 g_{2}^2 +\biggl(\frac{223}{80}g_{1}^2+\frac{135}{16} g_{2}^2 +16g_{3}^2\biggl)h_{t}^2  \biggl. \nonumber\\
&&-\biggl( \frac{43}{80}g_{1}^2- \frac{9}{16} g_{2}^2+ 16 g_{3}^2 \biggl) h_{b}^2+\frac{5}{2} Y_{4}(S) - 2 \lambda \biggl(3h_{t}^2+ h_{b}^2\biggl)+ \frac{3}{2} h_{t}^4-\frac{5}{4} h_{t}^2 h_{b}^2   \biggl. \nonumber\\
&&+\frac{11}{4} h_{b}^4+ Y_{2}(S)\biggl(\frac{5}{4}h_{b}^2 - \frac{9}{4}h_{t}^2\biggl)
- \eta _{4}(S)+\frac{3}{2}\lambda^2 \biggl],
\end{eqnarray}

\begin{eqnarray}{\label{eq:b}}
\frac{dh_b}{dt} &=&\frac{h_b}{16 \pi^2}\biggl(\frac{3}{2} h_{b}^2 - \frac{3}{2} h_{t}^2 +Y_{2}(S) - \sum_{\mathclap{i=1}}^{3} c_{i}^{'} g_{i}^2 \biggl)+\frac{h_b}{(16 \pi^2)^2}\biggl[ -\frac{127}{600}g_{1}^4-  \frac{23}{4} g_{2}^4 - 108 g_{3}^4 \biggl. \nonumber\\
&& \biggl. - \frac{27}{20} g_{1}^2 g_{2}^2+ \frac{31}{15} g_{1}^{2} g_{3}^2 +9 g_{3}^2 g_{2}^2 -\biggl (\frac{79}{80}g_{1}^2-\frac{9}{16} g_{2}^2 +16g_{3}^2\biggl)h_{t}^2 + \biggl( \frac{187}{80}g_{1}^2+ \frac{135}{16} g_{2}^2  \biggl. \nonumber\\
&&+ 16 g_{3}^2 \biggl) h_{b}^2 +\frac{5}{2} Y_{4}(S)- 2 \lambda \biggl(3h_{t}^2+3 h_{b}^2\biggl) + \frac{3}{2} h_{b}^4-\frac{5}{4} h_{t}^2 h_{b}^2 +\frac{11}{4} h_{t}^4  \biggl. \nonumber\\
 &&\biggl.+ Y_{2}(S)\biggl(\frac{5}{4}h_{t}^2 - \frac{9}{4}h_{b}^2\biggl) - \eta _{4}(S)+\frac{3}{2}\lambda^2 \biggl],
\end{eqnarray}

\begin{eqnarray}{\label{eq:tau}}
\frac{dh_\tau}{dt} &=&\frac{h_\tau}{16 \pi^2}\biggl(\frac{3}{2} h_{\tau}^2 + Y_{2}(S) - \sum_{\mathclap{i=1}}^{3} c_i^{''}g_{i}^2 \biggl)
+\frac{h_\tau}{(16 \pi^2)^2}\biggl[ \frac{1371}{200}g_{1}^4-  \frac{23}{4} g_{2}^4 -  \frac{27}{20} g_{1}^2 g_{2}^2 \biggl. \nonumber\\
&& \biggl.+\biggl(\frac{387}{80}g_{1}^2+ \frac{135}{16} g_{2}^2 \biggl)h_{\tau}^2 +\frac{5}{2} Y_{4}(S) - 6\lambda h_{t}^2+ \frac{3}{2}h_{\tau}^4 - \frac{9}{4}Y_{2}(S)h_{\tau}^2 \biggl. \nonumber\\
&&\biggl.- \eta _{4}(S) +\frac{3}{2} \lambda^2\biggl],
\end{eqnarray}

\begin{eqnarray}
\frac{d\lambda}{dt} &=&\frac{1}{16 \pi^2}\biggl[\frac{9}{4}\biggl(\frac{3}{25} g_{1}^4 + \frac{2}{5} g_{2}^2 g_{1}^2 + g_{2}^4 \biggl)-\biggl(\frac{9}{5} g_{1}^2 +9 g_{2}^2\biggl)\lambda 
+ 4 Y_{2}(S) \lambda -4H(S) +12 \lambda^2\biggl]
\biggl. \nonumber\\
&&\biggl.+\frac{1}{(16 \pi^2)^2}\biggl[ -78 \lambda^3 + 18\biggl(\frac{3}{5}g_{1}^2+ 3  g_{2}^2 \biggl)\lambda^2 +\biggl(- \frac{73}{8} g_{2}^4 + \frac{117}{20} g_{1}^{2} g_{2}^2 + \frac{1887}{200}g_{1}^4\biggl)\lambda\biggl. \nonumber\\
&&\biggl.+\frac{305}{8} g_{2}^6  - \frac{867}{120}g_{1}^2 g_{2}^4- \frac{1677}{200} g_{1}^4 g_{2}^2-\frac{3411}{1000} g_{1}^6 - 64 g_{3}^2\biggl(h_{t}^4+ h_{b}^4\biggl)- \frac{8}{5} g_{1}^2\biggl(2 h_{t}^4 - h_{b}^4 \biggl. \nonumber\\
&&\biggl.+3 h_{\tau}^4 \biggl)- \frac{3}{2}g_{2}^4 Y_{2}(S)+10 \lambda Y_{4}(S)+ \frac{3}{5}g_{1}^2(- \frac{57}{10}g_{1}^2 + 21 g_{2}^2)h_{t}^2+\biggl(\frac{3}{2}g_{1}^2 +9 g_{2}^2\biggl)h_{b}^2 \biggl. \nonumber\\
&&\biggl. +\biggl(-\frac{15}{2} g_{1}^2 +11 g_{2}^2\biggl)h_{\tau}^2 -24 \lambda^2 Y_{2}(S)-
 \lambda H(S)+6 \lambda h_{t}^2 h_{b}^2 +20 \biggl(3h_{t}^6 +3 h_{b}^6+h_{\tau}^6\biggl)\biggl. \nonumber\\
 &&\biggl. -12\biggl(h_{t}^4 h_{b}^2+h_{t}^2 h_{b}^4\biggl)\biggl],
\end{eqnarray}
where 
 $$ Y_{2}(S)=3 h_{t}^2+ 3 h_{b}^2+ h_{\tau}^2 ,$$
$$ Y_{4}(S)=\frac{1}{3}\biggl[3 \Sigma c_{i} g_{i}^2 h_{t}^2+3 \Sigma c_{i}^{'} g_{i}^2 h_{b}^2+ 3 \Sigma c_{i}^{''} g_{i}^2 h_{\tau}^2\biggl],$$

$$ H(S)=3 h_{t}^4+ 3 h_{b}^4+ h_{\tau}^4, $$

 $$\eta_{4}(S)=\frac{9}{4} \biggl[3 h_{t}^4 + 3 h_{b}^4 + h_{\tau}^4 -\frac{2}{3} h_{t}^2 h_{b}^2\biggl]$$

and $\displaystyle \lambda=  \frac{m_{h}^2}{v_0^2}$ is the Higgs self-coupling, $ m_h$ = 125.78 $\pm $ 0.26 GeV  is the Higgs mass \cite{higgs} and  $v_0$ = 174 GeV is the vacuum expectation value.
\\

The beta function coefficients for  NOn-SUSY case are given below:

$ \displaystyle c_{i}=
\left(\begin{array}{ccc}
0.85, & 2.25, & 8.00
\end{array}\right), $ 
$\displaystyle c_{i}^{'}=
\left(\begin{array}{ccc}
0.25 & 2.25, & 8.00  
\end{array}\right)$
and
$\displaystyle c_{i}^{''}=
\left(\begin{array}{ccc}
2.25 ,& 2.25 ,& 0.00  
\end{array}\right).$
\section*{APPENDIX B}

\subsubsection*{ RGEs for three neutrino mixing angles and phases \cite{rg2}: (neglecting higher order of $\theta_{13}$)}

\begin{equation}{\label{eq:t12}}
\dot{\theta}_{12}= -\frac{Ch_{\tau}^2}{32 \pi^2} \sin2\theta_{12} s_{23}^2 \frac{|m_1 e^{i\psi_1} +m_2 e^{i\psi_2}|^2}{\Delta{m_{21}^2}},
\end{equation}
$$
\dot{\theta}_{13}= \frac{Ch_{\tau}^2}{32 \pi^2}\sin2\theta_{12}\sin2\theta_{23}\frac{m_3}{\Delta{m_{31}^2}(1+\xi)}  
$$
\begin{equation}{\label{eq:t13}}
 \times \biggl[ m_1 \cos(\psi_{1}-\delta) -(1+\xi)m_2 \cos(\psi_2 - \delta)- \xi m_3 \cos\delta \biggl],
\end{equation}
\begin{equation}
\dot{\theta}_{23}= -\frac{C h_{\tau}^2}{32 \pi^2} \sin2\theta_{23}\frac{1}{\Delta m_{31}^2}\left[ c_{12}^2 |m_2 e^{i\psi_{2}} +m_3|^2 + s_{12}^2 \frac{|m_1 e^{i\psi_1} +m_3|^2}{1+\xi}\right],
\end{equation}

 where $\Delta m_{21}^2 = m_2^2 - m_1 ^2$ , $\Delta m_{31}^2 = m_3^2 - m_1 ^2$ and $\xi = \frac{\Delta m_{21}^2 }{\Delta m_{31}^2 }$. 
\subsubsection*{ RGEs for the three phases \cite{rg2}:}
 For  dirac  phase $\delta$:
\begin{equation}
 \dot{\delta} = \frac{C h_{\tau}^2}{32 \pi^2} \frac{\delta^{(-1)}}{\theta_{13}} + \frac{C h_{\tau}^2}{8 \pi^2} \delta^{(0)}, 
\end{equation}
where
\begin{eqnarray}
\delta^{(-1)}& = &\sin 2 \theta_{12} \sin 2 \theta_{23}\frac{m_3}{\Delta m_{31}^2(1+\xi) } \times \biggl[ m_1 \sin(\psi_1- \delta) 
\biggl. \nonumber\\
&&\biggl.-(1+\xi) m_2 \sin(\psi_2 - \delta) + \xi m_3 \sin\delta\biggl], 
\end{eqnarray}

\begin{eqnarray}
\delta^{(0)} &=& \frac{m_1 m_2 s_{23}^2 \sin(\psi_1 - \psi_2)}{\Delta m_{21}^2} \biggl. \nonumber\\
&&\biggl.  +  m_3 s_{12}^2\biggl[ \frac{m_1 \cos 2\theta_{23} \sin\psi_1}{\Delta m_{31}^2(1+\xi)}+ \frac{m_2 c_{23}^2 \sin(2\delta - \psi_2)}{\Delta m_{31}^2} \biggl]\biggl. \nonumber\\
&&\biggl. +  m_3 c_{12}^2\biggl[ \frac{m_1 c_{23}^2 \sin(2 \delta - \psi_1)}{\Delta m_{31}^2(1+\xi)}+ \frac{m_2 \cos(2\theta_{23})\sin\psi_2}{\Delta m_{31}^2} \biggl].
\end{eqnarray}

For Majorana phase $\psi_1$ \cite{rg2}:
\begin{eqnarray}
\dot{\psi_1}&=&\frac{C h_{\tau}^2}{8 \pi^2} \left[ m_3 \cos 2 \theta_{23} \frac{m_1 s_{12}^2 \sin \psi_1 + (1+ \xi) m_2 c_{12}^2 \sin \psi_2}{\Delta m_{31}^2(1+ \xi)}\right]\biggl. \nonumber\\
 &&\biggl. + \frac{C h_{\tau}^2}{8 \pi^2}\left[ \frac{m_1 m_2 c_{12}^2 s_{23}^2 sin(\psi_1 - \psi_2)}{
\Delta m_{21}^2}\right] 
\end{eqnarray}
 For Majorana phase $\psi_2$:
\begin{eqnarray}
 \dot{\psi_2}&=&\frac{C h_{\tau}^2}{8 \pi^2}\left[  m_3 \cos 2 \theta_{23} \frac{m_1 s_{12}^2 \sin \psi_1 +(1+\xi) m_2 c_{12}^2 \sin \psi_2  }{\Delta m_{31}^2 (1+\xi)}\right]\biggl. \nonumber\\
 &&\biggl.+\frac{C h_{\tau}^2}{8 \pi^2}\left[ \frac{m_1 m_2 s_{12}^2 s_{23}^2 \sin(\psi_1 - \psi_2
)}{\Delta m_{21}^2} \right] 
\end{eqnarray}
\subsubsection*{RGEs for neutrino mass eigenvalues \cite{rg2}:}
\begin{equation}
\dot{m_1} = \frac{1}{16 \pi^2}\left[  \alpha + C h_{\tau}^2 (2 s_{12}^2 s_{23}^2 +F_1)\right]m_1,
\end{equation}
\begin{equation}
\dot{m_2} = \frac{1}{16 \pi^2}\left[  \alpha + C h_{\tau}^2 (2 c_{12}^2 s_{23}^2 +F_2)\right]m_2,
\end{equation}
\begin{equation}
\dot{m_3} = \frac{1}{16 \pi^2}\left[  \alpha + 2C h_{\tau}^2 c_{13}^2 c_{23} \right]m_3,
\end{equation}
where
\begin{equation}
 F_1 = - s_{13}\sin 2 \theta_{12} \sin 2 \theta_{23} \cos \delta + 2 s_{13}^2 c_{12}^2 c_{23}^2,
 \end{equation}
 \begin{equation}
 F_2 =  s_{13}\sin 2 \theta_{12} \sin 2 \theta_{23} \cos \delta + 2 s_{13}^2 s_{12}^2 s_{23}^2.
 \end{equation}
  
For MSSM case:
  
$$\alpha= -\frac{6}{5} g_{1}^2 - 6 g_2 ^2 +6 h_t^2 $$ and $$C=1.$$

For SM case: 
 
$$\alpha= -3 g_{2}^2 + 2 h_{\tau}^2  +6 h_{t}^2 +6 h_{b}^2 + \lambda,$$
$$C=-\frac{3}{2}$$ and $\lambda$ is the Higgs self-coupling in the SM.

\section*{Conflict of Interest}
The author(s) declare(s) that there is no conflict of interest regarding the publication of this paper.
\section*{Funding Statement}
This study is funded by $SCOAP^3$.
\section*{Acknowledgements}
One of the author (KHD) would like to thank $SCOAP^3$ for  financial support.

\bibliographystyle{ieeetr}
\bibliography{tbmnew1}

\begin{thebibliography}{10}

\bibitem{c2}
M.~C. Gonzalez-Garcia, M.~Maltoni, and T.~Schwetz, ``{NuFIT: Three-Flavour
  Global Analyses of Neutrino Oscillation Experiments},'' {\em Universe},
  vol.~7, no.~12, p.~459, 2021.

\bibitem{tbm}
P.~F. Harrison, D.~H. Perkins, and W.~G. Scott, ``{Tri-bimaximal mixing and the
  neutrino oscillation data},'' {\em Phys. Lett. B}, vol.~530, p.~167, 2002.

\bibitem{tk}
K.~Abe, N.~Abgrall, Y.~Ajima, H.~Aihara, J.~B. Albert, C.~Andreopoulos,
  B.~Andrieu, S.~Aoki, O.~Araoka, J.~Argyriades, A.~Ariga, T.~Ariga,
  S.~Assylbekov, D.~Autiero, A.~Badertscher, M.~Barbi, G.~J. Barker, G.~Barr,
  M.~Bass, F.~Bay, S.~Bentham, V.~Berardi, B.~E. Berger, I.~Bertram,
  M.~Besnier, J.~Beucher, D.~Beznosko, S.~Bhadra, F.~d.~M. Blaszczyk,
  A.~Blondel, C.~Bojechko, J.~Bouchez, S.~B. Boyd, A.~Bravar, C.~Bronner, D.~G.
  Brook-Roberge, N.~Buchanan, H.~Budd, D.~Calvet, S.~L. Cartwright, A.~Carver,
  R.~Castillo, M.~G. Catanesi, A.~Cazes, A.~Cervera, C.~Chavez, S.~Choi,
  G.~Christodoulou, J.~Coleman, W.~Coleman, G.~Collazuol, K.~Connolly,
  A.~Curioni, A.~Dabrowska, I.~Danko, R.~Das, G.~S. Davies, S.~Davis, M.~Day,
  G.~De~Rosa, J.~P. A.~M. de~Andr\'e, P.~de~Perio, A.~Delbart, C.~Densham,
  F.~Di~Lodovico, S.~Di~Luise, P.~Dinh~Tran, J.~Dobson, U.~Dore, O.~Drapier,
  F.~Dufour, J.~Dumarchez, S.~Dytman, M.~Dziewiecki, M.~Dziomba, S.~Emery,
  A.~Ereditato, L.~Escudero, L.~S. Esposito, M.~Fechner, A.~Ferrero, A.~J.
  Finch, E.~Frank, Y.~Fujii, Y.~Fukuda, V.~Galymov, F.~C. Gannaway, A.~Gaudin,
  A.~Gendotti, M.~A. George, S.~Giffin, C.~Giganti, K.~Gilje, T.~Golan,
  M.~Goldhaber, J.~J. Gomez-Cadenas, M.~Gonin, N.~Grant, A.~Grant,
  P.~Gumplinger, P.~Guzowski, A.~Haesler, M.~D. Haigh, K.~Hamano, C.~Hansen,
  D.~Hansen, T.~Hara, P.~F. Harrison, B.~Hartfiel, M.~Hartz, T.~Haruyama,
  T.~Hasegawa, N.~C. Hastings, S.~Hastings, A.~Hatzikoutelis, K.~Hayashi,
  Y.~Hayato, C.~Hearty, R.~L. Helmer, R.~Henderson, N.~Higashi, J.~Hignight,
  E.~Hirose, J.~Holeczek, S.~Horikawa, A.~Hyndman, A.~K. Ichikawa, K.~Ieki,
  M.~Ieva, M.~Iida, M.~Ikeda, J.~Ilic, J.~Imber, T.~Ishida, C.~Ishihara,
  T.~Ishii, S.~J. Ives, M.~Iwasaki, K.~Iyogi, A.~Izmaylov, B.~Jamieson, R.~A.
  Johnson, K.~K. Joo, G.~V. Jover-Manas, C.~K. Jung, H.~Kaji, T.~Kajita,
  H.~Kakuno, J.~Kameda, K.~Kaneyuki, D.~Karlen, K.~Kasami, I.~Kato, E.~Kearns,
  M.~Khabibullin, F.~Khanam, A.~Khotjantsev, D.~Kielczewska, T.~Kikawa, J.~Kim,
  J.~Y. Kim, S.~B. Kim, N.~Kimura, B.~Kirby, J.~Kisiel, P.~Kitching,
  T.~Kobayashi, G.~Kogan, S.~Koike, A.~Konaka, L.~L. Kormos, A.~Korzenev,
  K.~Koseki, Y.~Koshio, Y.~Kouzuma, K.~Kowalik, V.~Kravtsov, I.~Kreslo,
  W.~Kropp, H.~Kubo, Y.~Kudenko, N.~Kulkarni, R.~Kurjata, T.~Kutter, J.~Lagoda,
  K.~Laihem, M.~Laveder, K.~P. Lee, P.~T. Le, J.~M. Levy, C.~Licciardi, I.~T.
  Lim, T.~Lindner, R.~P. Litchfield, M.~Litos, A.~Longhin, G.~D. Lopez, P.~F.
  Loverre, L.~Ludovici, T.~Lux, M.~Macaire, K.~Mahn, Y.~Makida, M.~Malek,
  S.~Manly, A.~Marchionni, A.~D. Marino, J.~Marteau, J.~F. Martin, T.~Maruyama,
  T.~Maryon, J.~Marzec, P.~Masliah, E.~L. Mathie, C.~Matsumura, K.~Matsuoka,
  V.~Matveev, K.~Mavrokoridis, E.~Mazzucato, N.~McCauley, K.~S. McFarland,
  C.~McGrew, T.~McLachlan, M.~Messina, W.~Metcalf, C.~Metelko, M.~Mezzetto,
  P.~Mijakowski, C.~A. Miller, A.~Minamino, O.~Mineev, S.~Mine, A.~D. Missert,
  G.~Mituka, M.~Miura, K.~Mizouchi, L.~Monfregola, F.~Moreau, B.~Morgan,
  S.~Moriyama, A.~Muir, A.~Murakami, M.~Murdoch, S.~Murphy, J.~Myslik,
  T.~Nakadaira, M.~Nakahata, T.~Nakai, K.~Nakajima, T.~Nakamoto, K.~Nakamura,
  S.~Nakayama, T.~Nakaya, D.~Naples, M.~L. Navin, B.~Nelson, T.~C. Nicholls,
  K.~Nishikawa, H.~Nishino, J.~A. Nowak, M.~Noy, Y.~Obayashi, T.~Ogitsu,
  H.~Ohhata, T.~Okamura, K.~Okumura, T.~Okusawa, S.~M. Oser, M.~Otani, R.~A.
  Owen, Y.~Oyama, T.~Ozaki, M.~Y. Pac, V.~Palladino, V.~Paolone, P.~Paul,
  D.~Payne, G.~F. Pearce, J.~D. Perkin, V.~Pettinacci, F.~Pierre, E.~Poplawska,
  B.~Popov, M.~Posiadala, J.-M. Poutissou, R.~Poutissou, P.~Przewlocki,
  W.~Qian, J.~L. Raaf, E.~Radicioni, P.~N. Ratoff, T.~M. Raufer, M.~Ravonel,
  M.~Raymond, F.~Retiere, A.~Robert, P.~A. Rodrigues, E.~Rondio, J.~M. Roney,
  B.~Rossi, S.~Roth, A.~Rubbia, D.~Ruterbories, S.~Sabouri, R.~Sacco,
  K.~Sakashita, F.~S\'anchez, A.~Sarrat, K.~Sasaki, K.~Scholberg, J.~Schwehr,
  M.~Scott, D.~I. Scully, Y.~Seiya, T.~Sekiguchi, H.~Sekiya, M.~Shibata,
  Y.~Shimizu, M.~Shiozawa, S.~Short, M.~Siyad, R.~J. Smith, M.~Smy, J.~T.
  Sobczyk, H.~Sobel, M.~Sorel, A.~Stahl, P.~Stamoulis, J.~Steinmann, B.~Still,
  J.~Stone, C.~Strabel, L.~R. Sulak, R.~Sulej, P.~Sutcliffe, A.~Suzuki,
  K.~Suzuki, S.~Suzuki, S.~Y. Suzuki, Y.~Suzuki, Y.~Suzuki, T.~Szeglowski,
  M.~Szeptycka, R.~Tacik, M.~Tada, S.~Takahashi, A.~Takeda, Y.~Takenaga,
  Y.~Takeuchi, K.~Tanaka, H.~A. Tanaka, M.~Tanaka, M.~M. Tanaka, N.~Tanimoto,
  K.~Tashiro, I.~Taylor, A.~Terashima, D.~Terhorst, R.~Terri, L.~F. Thompson,
  A.~Thorley, W.~Toki, T.~Tomaru, Y.~Totsuka, C.~Touramanis, T.~Tsukamoto,
  M.~Tzanov, Y.~Uchida, K.~Ueno, A.~Vacheret, M.~Vagins, G.~Vasseur,
  T.~Wachala, J.~J. Walding, A.~V. Waldron, C.~W. Walter, P.~J. Wanderer,
  J.~Wang, M.~A. Ward, G.~P. Ward, D.~Wark, M.~O. Wascko, A.~Weber, R.~Wendell,
  N.~West, L.~H. Whitehead, G.~Wikstr\"om, R.~J. Wilkes, M.~J. Wilking, J.~R.
  Wilson, R.~J. Wilson, T.~Wongjirad, S.~Yamada, Y.~Yamada, A.~Yamamoto,
  K.~Yamamoto, Y.~Yamanoi, H.~Yamaoka, C.~Yanagisawa, T.~Yano, S.~Yen,
  N.~Yershov, M.~Yokoyama, A.~Zalewska, J.~Zalipska, L.~Zambelli, K.~Zaremba,
  M.~Ziembicki, E.~D. Zimmerman, M.~Zito, and J.~\ifmmode~\dot{Z}\else
  \.{Z}\fi{}muda, ``{Indication of Electron Neutrino Appearance from an
  Accelerator-produced Off-axis Muon Neutrino Beam},'' {\em Phys. Rev. Lett.},
  vol.~107, p.~041801, 2011.

\bibitem{mnos}
P.~Adamson, D.~J. Auty, D.~S. Ayres, C.~Backhouse, G.~Barr, M.~Betancourt,
  M.~Bishai, A.~Blake, G.~J. Bock, D.~J. Boehnlein, D.~Bogert, S.~V. Cao,
  S.~Cavanaugh, D.~Cherdack, S.~Childress, J.~A.~B. Coelho, L.~Corwin,
  D.~Cronin-Hennessy, I.~Z. Danko, J.~K. de~Jong, N.~E. Devenish, M.~V. Diwan,
  M.~Dorman, C.~O. Escobar, J.~J. Evans, E.~Falk, G.~J. Feldman, M.~V. Frohne,
  H.~R. Gallagher, R.~A. Gomes, M.~C. Goodman, P.~Gouffon, N.~Graf, R.~Gran,
  K.~Grzelak, A.~Habig, J.~Hartnell, R.~Hatcher, A.~Himmel, A.~Holin, X.~Huang,
  J.~Hylen, G.~M. Irwin, Z.~Isvan, D.~E. Jaffe, C.~James, D.~Jensen, T.~Kafka,
  S.~M.~S. Kasahara, G.~Koizumi, S.~Kopp, M.~Kordosky, A.~Kreymer, K.~Lang,
  G.~Lefeuvre, J.~Ling, P.~J. Litchfield, L.~Loiacono, P.~Lucas, W.~A. Mann,
  M.~L. Marshak, M.~Mathis, N.~Mayer, A.~M. McGowan, R.~Mehdiyev, J.~R. Meier,
  M.~D. Messier, D.~G. Michael, W.~H. Miller, S.~R. Mishra, J.~Mitchell, C.~D.
  Moore, L.~Mualem, S.~Mufson, J.~Musser, D.~Naples, J.~K. Nelson, H.~B.
  Newman, R.~J. Nichol, J.~A. Nowak, J.~P. Ochoa-Ricoux, W.~P. Oliver,
  M.~Orchanian, J.~Paley, R.~B. Patterson, G.~Pawloski, G.~F. Pearce,
  S.~Phan-Budd, R.~K. Plunkett, X.~Qiu, J.~Ratchford, B.~Rebel, C.~Rosenfeld,
  H.~A. Rubin, M.~C. Sanchez, J.~Schneps, A.~Schreckenberger, P.~Schreiner,
  P.~Shanahan, R.~Sharma, A.~Sousa, N.~Tagg, R.~L. Talaga, J.~Thomas, M.~A.
  Thomson, R.~Toner, D.~Torretta, G.~Tzanakos, J.~Urheim, P.~Vahle, B.~Viren,
  J.~J. Walding, A.~Weber, R.~C. Webb, C.~White, L.~Whitehead, S.~G. Wojcicki,
  T.~Yang, and R.~Zwaska, ``{Improved search for muon-neutrino to
  electron-neutrino oscillations in MINOS},'' {\em Phys. Rev. Lett.}, vol.~107,
  p.~181802, 2011.

\bibitem{dcz}
Y.~Abe, C.~Aberle, T.~Akiri, J.~C. dos Anjos, F.~Ardellier, A.~F. Barbosa,
  A.~Baxter, M.~Bergevin, A.~Bernstein, T.~J.~C. Bezerra, L.~Bezrukhov,
  E.~Blucher, M.~Bongrand, N.~S. Bowden, C.~Buck, J.~Busenitz, A.~Cabrera,
  E.~Caden, L.~Camilleri, R.~Carr, M.~Cerrada, P.-J. Chang, P.~Chimenti,
  T.~Classen, A.~P. Collin, E.~Conover, J.~M. Conrad, S.~Cormon, J.~I.
  Crespo-Anad\'on, M.~Cribier, K.~Crum, A.~Cucoanes, M.~V. D'Agostino,
  E.~Damon, J.~V. Dawson, S.~Dazeley, M.~Dierckxsens, D.~Dietrich, Z.~Djurcic,
  M.~Dracos, V.~Durand, Y.~Efremenko, M.~Elnimr, Y.~Endo, A.~Etenko, E.~Falk,
  M.~Fallot, M.~Fechner, F.~von Feilitzsch, J.~Felde, S.~M. Fernandes,
  D.~Franco, A.~J. Franke, M.~Franke, H.~Furuta, R.~Gama, I.~Gil-Botella,
  L.~Giot, M.~G\"oger-Neff, L.~F.~G. Gonzalez, M.~C. Goodman, J.~T. Goon,
  D.~Greiner, B.~Guillon, N.~Haag, C.~Hagner, T.~Hara, F.~X. Hartmann,
  J.~Hartnell, T.~Haruna, J.~Haser, A.~Hatzikoutelis, T.~Hayakawa, M.~Hofmann,
  G.~A. Horton-Smith, M.~Ishitsuka, J.~Jochum, C.~Jollet, C.~L. Jones,
  F.~Kaether, L.~Kalousis, Y.~Kamyshkov, D.~M. Kaplan, T.~Kawasaki, G.~Keefer,
  E.~Kemp, H.~de~Kerret, Y.~Kibe, T.~Konno, D.~Kryn, M.~Kuze, T.~Lachenmaier,
  C.~E. Lane, C.~Langbrandtner, T.~Lasserre, A.~Letourneau, D.~Lhuillier, H.~P.
  Lima, M.~Lindner, Y.~Liu, J.~M. L\'opez-Castan\~o, J.~M. LoSecco, B.~K.
  Lubsandorzhiev, S.~Lucht, D.~McKee, J.~Maeda, C.~N. Maesano, C.~Mariani,
  J.~Maricic, J.~Martino, T.~Matsubara, G.~Mention, A.~Meregaglia, T.~Miletic,
  R.~Milincic, A.~Milzstajn, H.~Miyata, D.~Motta, T.~A. Mueller, Y.~Nagasaka,
  K.~Nakajima, P.~Novella, M.~Obolensky, L.~Oberauer, A.~Onillon, A.~Osborn,
  I.~Ostrovskiy, C.~Palomares, S.~J.~M. Peeters, I.~M. Pepe, S.~Perasso,
  P.~Perrin, P.~Pfahler, A.~Porta, W.~Potzel, R.~Queval, J.~Reichenbacher,
  B.~Reinhold, A.~Remoto, D.~Reyna, M.~R\"ohling, S.~Roth, H.~A. Rubin,
  Y.~Sakamoto, R.~Santorelli, F.~Sato, S.~Sch\"onert, S.~Schoppmann, U.~Schwan,
  T.~Schwetz, M.~H. Shaevitz, D.~Shrestha, J.-L. Sida, V.~Sinev,
  M.~Skorokhvatov, E.~Smith, J.~Spitz, A.~Stahl, I.~Stancu, M.~Strait,
  A.~St\"uken, F.~Suekane, S.~Sukhotin, T.~Sumiyoshi, Y.~Sun, Z.~Sun,
  R.~Svoboda, H.~Tabata, N.~Tamura, K.~Terao, A.~Tonazzo, M.~Toups, H.~H.
  Trinh~Thi, C.~Veyssiere, S.~Wagner, H.~Watanabe, B.~White, C.~Wiebusch,
  L.~Winslow, M.~Worcester, M.~Wurm, E.~Yanovitch, F.~Yermia, K.~Zbiri, and
  V.~Zimmer, ``{Indication of Reactor $\bar{\nu}_e$ Disappearance in the Double
  Chooz Experiment},'' {\em Phys. Rev. Lett.}, vol.~108, p.~131801, 2012.

\bibitem{db}
F.~P. An, J.~Z. Bai, A.~B. Balantekin, H.~R. Band, D.~Beavis, W.~Beriguete,
  M.~Bishai, S.~Blyth, K.~Boddy, R.~L. Brown, B.~Cai, G.~F. Cao, J.~Cao,
  R.~Carr, W.~T. Chan, J.~F. Chang, Y.~Chang, C.~Chasman, H.~S. Chen, H.~Y.
  Chen, S.~J. Chen, S.~M. Chen, X.~C. Chen, X.~H. Chen, X.~S. Chen, Y.~Chen,
  Y.~X. Chen, J.~J. Cherwinka, M.~C. Chu, J.~P. Cummings, Z.~Y. Deng, Y.~Y.
  Ding, M.~V. Diwan, L.~Dong, E.~Draeger, X.~F. Du, D.~A. Dwyer, W.~R. Edwards,
  S.~R. Ely, S.~D. Fang, J.~Y. Fu, Z.~W. Fu, L.~Q. Ge, V.~Ghazikhanian, R.~L.
  Gill, J.~Goett, M.~Gonchar, G.~H. Gong, H.~Gong, Y.~A. Gornushkin, L.~S.
  Greenler, W.~Q. Gu, M.~Y. Guan, X.~H. Guo, R.~W. Hackenburg, R.~L. Hahn,
  S.~Hans, M.~He, Q.~He, W.~S. He, K.~M. Heeger, Y.~K. Heng, P.~Hinrichs, T.~H.
  Ho, Y.~K. Hor, Y.~B. Hsiung, B.~Z. Hu, T.~Hu, T.~Hu, H.~X. Huang, H.~Z.
  Huang, P.~W. Huang, X.~Huang, X.~T. Huang, P.~Huber, Z.~Isvan, D.~E. Jaffe,
  S.~Jetter, X.~L. Ji, X.~P. Ji, H.~J. Jiang, W.~Q. Jiang, J.~B. Jiao, R.~A.
  Johnson, L.~Kang, S.~H. Kettell, M.~Kramer, K.~K. Kwan, M.~W. Kwok, T.~Kwok,
  C.~Y. Lai, W.~C. Lai, W.~H. Lai, K.~Lau, L.~Lebanowski, J.~Lee, M.~K.~P. Lee,
  R.~Leitner, J.~K.~C. Leung, K.~Y. Leung, C.~A. Lewis, B.~Li, F.~Li, G.~S. Li,
  J.~Li, Q.~J. Li, S.~F. Li, W.~D. Li, X.~B. Li, X.~N. Li, X.~Q. Li, Y.~Li,
  Z.~B. Li, H.~Liang, J.~Liang, C.~J. Lin, G.~L. Lin, S.~K. Lin, S.~X. Lin,
  Y.~C. Lin, J.~J. Ling, J.~M. Link, L.~Littenberg, B.~R. Littlejohn, B.~J.
  Liu, C.~Liu, D.~W. Liu, H.~Liu, J.~C. Liu, J.~L. Liu, S.~Liu, X.~Liu, Y.~B.
  Liu, C.~Lu, H.~Q. Lu, A.~Luk, K.~B. Luk, T.~Luo, X.~L. Luo, L.~H. Ma, Q.~M.
  Ma, X.~B. Ma, X.~Y. Ma, Y.~Q. Ma, B.~Mayes, K.~T. McDonald, M.~C. McFarlane,
  R.~D. McKeown, Y.~Meng, D.~Mohapatra, J.~E. Morgan, Y.~Nakajima,
  J.~Napolitano, D.~Naumov, I.~Nemchenok, C.~Newsom, H.~Y. Ngai, W.~K. Ngai,
  Y.~B. Nie, Z.~Ning, J.~P. Ochoa-Ricoux, D.~Oh, A.~Olshevski, A.~Pagac,
  S.~Patton, C.~Pearson, V.~Pec, J.~C. Peng, L.~E. Piilonen, L.~Pinsky,
  C.~S.~J. Pun, F.~Z. Qi, M.~Qi, X.~Qian, N.~Raper, R.~Rosero, B.~Roskovec,
  X.~C. Ruan, B.~Seilhan, B.~B. Shao, K.~Shih, H.~Steiner, P.~Stoler, G.~X.
  Sun, J.~L. Sun, Y.~H. Tam, H.~K. Tanaka, X.~Tang, H.~Themann, Y.~Torun,
  S.~Trentalange, O.~Tsai, K.~V. Tsang, R.~H.~M. Tsang, C.~Tull, B.~Viren,
  S.~Virostek, V.~Vorobel, C.~H. Wang, L.~S. Wang, L.~Y. Wang, L.~Z. Wang,
  M.~Wang, N.~Y. Wang, R.~G. Wang, T.~Wang, W.~Wang, X.~Wang, X.~Wang, Y.~F.
  Wang, Z.~Wang, Z.~Wang, Z.~M. Wang, D.~M. Webber, Y.~D. Wei, L.~J. Wen, D.~L.
  Wenman, K.~Whisnant, C.~G. White, L.~Whitehead, C.~A. Whitten, J.~Wilhelmi,
  T.~Wise, H.~C. Wong, H.~L.~H. Wong, J.~Wong, E.~T. Worcester, F.~F. Wu,
  Q.~Wu, D.~M. Xia, S.~T. Xiang, Q.~Xiao, Z.~Z. Xing, G.~Xu, J.~Xu, J.~Xu,
  J.~L. Xu, W.~Xu, Y.~Xu, T.~Xue, C.~G. Yang, L.~Yang, M.~Ye, M.~Yeh, Y.~S.
  Yeh, K.~Yip, B.~L. Young, Z.~Y. Yu, L.~Zhan, C.~Zhang, F.~H. Zhang, J.~W.
  Zhang, Q.~M. Zhang, K.~Zhang, Q.~X. Zhang, S.~H. Zhang, Y.~C. Zhang, Y.~H.
  Zhang, Y.~X. Zhang, Z.~J. Zhang, Z.~P. Zhang, Z.~Y. Zhang, J.~Zhao, Q.~W.
  Zhao, Y.~B. Zhao, L.~Zheng, W.~L. Zhong, L.~Zhou, Z.~Y. Zhou, H.~L. Zhuang,
  and J.~H. Zou, ``{Observation of electron-antineutrino disappearance at Daya
  Bay},'' {\em Phys. Rev. Lett.}, vol.~108, p.~171803, 2012.

\bibitem{reno}
J.~K. Ahn, S.~Chebotaryov, J.~H. Choi, S.~Choi, W.~Choi, Y.~Choi, H.~I. Jang,
  J.~S. Jang, E.~J. Jeon, I.~S. Jeong, K.~K. Joo, B.~R. Kim, B.~C. Kim, H.~S.
  Kim, J.~Y. Kim, S.~B. Kim, S.~H. Kim, S.~Y. Kim, W.~Kim, Y.~D. Kim, J.~Lee,
  J.~K. Lee, I.~T. Lim, K.~J. Ma, M.~Y. Pac, I.~G. Park, J.~S. Park, K.~S.
  Park, J.~W. Shin, K.~Siyeon, B.~S. Yang, I.~S. Yeo, S.~H. Yi, and I.~Yu,
  ``Observation of reactor electron antineutrinos disappearance in the reno
  experiment,'' {\em Phys. Rev. Lett.}, vol.~108, p.~191802, 2012.

\bibitem{lia}
I.~Gogoladze, F.~Nasir, and Q.~Shafi, ``{Non-Universal Gaugino Masses and
  Natural Supersymmetry},'' {\em Int. J. Mod. Phys. A}, vol.~28, p.~1350046,
  2013.

\bibitem{lib}
J.~Fan, M.~Reece, and L.-T. Wang, ``{Precision Natural SUSY at CEPC, FCC-ee,
  and ILC},'' {\em JHEP}, vol.~08, p.~152, 2015.

\bibitem{ntbm1}
C.~H. Albright and W.~Rodejohann, ``{Comparing Trimaximal Mixing and Its
  Variants with Deviations from Tri-bimaximal Mixing},'' {\em Eur. Phys. J. C},
  vol.~62, pp.~599--608, 2009.

\bibitem{ntbm2}
X.-G. He and A.~Zee, ``{Minimal Modification to Tri-bimaximal Mixing},'' {\em
  Phys. Rev. D}, vol.~84, p.~053004, 2011.

\bibitem{hel}
K.~H. Devi, K.~S. Singh, and N.~N. Singh, ``{Effects of Variations of SUSY
  Breaking Scale on Neutrino Parameters at Low Energy Scale under Radiative
  Corrections},'' {\em Adv. High Energy Phys.}, vol.~2022, p.~5780441, 2022.

\bibitem{rgg1}
P.~H. Chankowski and Z.~Pluciennik, ``{Renormalization group equations for
  seesaw neutrino masses},'' {\em Phys. Lett. B}, vol.~316, pp.~312--317, 1993.

\bibitem{rgg2}
K.~S. Babu, C.~N. Leung, and J.~T. Pantaleone, ``{Renormalization of the
  neutrino mass operator},'' {\em Phys. Lett. B}, vol.~319, pp.~191--198, 1993.

\bibitem{rgg3}
S.~Antusch, M.~Drees, J.~Kersten, M.~Lindner, and M.~Ratz, ``{Neutrino mass
  operator renormalization revisited},'' {\em Phys. Lett. B}, vol.~519,
  pp.~238--242, 2001.

\bibitem{rgg4}
S.~Antusch, J.~Kersten, M.~Lindner, M.~Ratz, and M.~A. Schmidt, ``{Running
  neutrino mass parameters in see-saw scenarios},'' {\em JHEP}, vol.~03,
  p.~024, 2005.

\bibitem{rgg5}
M.~Chala and A.~Titov, ``{Neutrino masses in the Standard Model effective field
  theory},'' {\em Phys. Rev. D}, vol.~104, no.~3, p.~035002, 2021.

\bibitem{da}
K.~S. Singh and N.~N. Singh, ``{Effects of the Variation of SUSY Breaking Scale
  on Yukawa and Gauge Couplings Unification},'' {\em Adv. High Energy Phys.},
  vol.~2015, p.~652029, 2015.

\bibitem{dac}
K.~S. Singh, S.~Roy, and N.~N. Singh, ``{Stability of neutrino parameters and
  self-complementarity relation with varying SUSY breaking scale},'' {\em Phys.
  Rev. D}, vol.~97, no.~5, p.~055038, 2018.

\bibitem{Kang:2014mka}
S.~K. Kang and C.~S. Kim, ``{Prediction of leptonic CP phase from
  perturbatively modified tribimaximal (or bimaximal) mixing},'' {\em Phys.
  Rev. D}, vol.~90, no.~7, p.~077301, 2014.

\bibitem{Rodejohann:2011uz}
W.~Rodejohann, H.~Zhang, and S.~Zhou, ``{Systematic search for successful
  lepton mixing patterns with nonzero $\theta_{13}$},'' {\em Nucl. Phys. B},
  vol.~855, pp.~592--607, 2012.

\bibitem{He:2011gb}
X.-G. He and A.~Zee, ``{Minimal Modification to Tri-bimaximal Mixing},'' {\em
  Phys. Rev. D}, vol.~84, p.~053004, 2011.

\bibitem{Albright:2008rp}
C.~H. Albright and W.~Rodejohann, ``{Comparing Trimaximal Mixing and Its
  Variants with Deviations from Tri-bimaximal Mixing},'' {\em Eur. Phys. J. C},
  vol.~62, pp.~599--608, 2009.

\bibitem{Garg:2018jsg}
S.~K. Garg, ``{Model independent analysis of Dirac CP violating phase for some
  well-known mixing scenarios},'' {\em Int. J. Mod. Phys. A}, vol.~36, no.~18,
  p.~2150118, 2021.

\bibitem{ntbm}
S.-P. Li, Yuan-Yuan-Li, X.-S. Yan, and X.~Zhang, ``{Next-to-tribimaximal mixing
  against CP violation and baryon asymmetry signs},'' {\em Phys. Rev. D},
  vol.~105, p.~9, 2022.

\bibitem{rg1}
V.~D. Barger, M.~S. Berger, and P.~Ohmann, ``{Supersymmetric grand unified
  theories: Two loop evolution of gauge and Yukawa couplings},'' {\em Phys.
  Rev. D}, vol.~47, pp.~1093--1113, 1993.

\bibitem{beta}
D.~R.~T. Jones and L.~Mezincescu, ``{The Beta Function in Supersymmetric
  {Yang-Mills} Theory},'' {\em Phys. Lett. B}, vol.~136, pp.~242--244, 1984.

\bibitem{pgd}
P.~D. Group, R.~L. Workman, V.~D. Burkert, V.~Crede, E.~Klempt, U.~Thoma,
  L.~Tiator, K.~Agashe, G.~Aielli, B.~C. Allanach, C.~Amsler, M.~Antonelli,
  E.~C. Aschenauer, D.~M. Asner, H.~Baer, S.~Banerjee, R.~M. Barnett,
  L.~Baudis, C.~W. Bauer, J.~J. Beatty, V.~I. Belousov, J.~Beringer,
  A.~Bettini, O.~Biebel, K.~M. Black, E.~Blucher, R.~Bonventre, V.~V.
  Bryzgalov, O.~Buchmuller, M.~A. Bychkov, R.~N. Cahn, M.~Carena, A.~Ceccucci,
  A.~Cerri, R.~S. Chivukula, G.~Cowan, K.~Cranmer, O.~Cremonesi, G.~D'Ambrosio,
  T.~Damour, D.~de~Florian, A.~de~Gouvêa, T.~DeGrand, P.~de~Jong, S.~Demers,
  B.~A. Dobrescu, M.~D'Onofrio, M.~Doser, H.~K. Dreiner, P.~Eerola, U.~Egede,
  S.~Eidelman, A.~X. El-Khadra, J.~Ellis, S.~C. Eno, J.~Erler, V.~V. Ezhela,
  W.~Fetscher, B.~D. Fields, A.~Freitas, H.~Gallagher, Y.~Gershtein,
  T.~Gherghetta, M.~C. Gonzalez-Garcia, M.~Goodman, C.~Grab, A.~V. Gritsan,
  C.~Grojean, D.~E. Groom, M.~Grünewald, A.~Gurtu, T.~Gutsche, H.~E. Haber,
  M.~Hamel, C.~Hanhart, S.~Hashimoto, Y.~Hayato, A.~Hebecker, S.~Heinemeyer,
  J.~J. Hernández-Rey, K.~Hikasa, J.~Hisano, A.~Höcker, J.~Holder, L.~Hsu,
  J.~Huston, T.~Hyodo, A.~Ianni, M.~Kado, M.~Karliner, U.~F. Katz, M.~Kenzie,
  V.~A. Khoze, S.~R. Klein, F.~Krauss, M.~Kreps, P.~Križan, B.~Krusche,
  Y.~Kwon, O.~Lahav, J.~Laiho, L.~P. Lellouch, J.~Lesgourgues, A.~R. Liddle,
  Z.~Ligeti, C.-J. Lin, C.~Lippmann, T.~M. Liss, L.~Littenberg, C.~Lourenço,
  K.~S. Lugovsky, S.~B. Lugovsky, A.~Lusiani, Y.~Makida, F.~Maltoni, T.~Mannel,
  A.~V. Manohar, W.~J. Marciano, A.~Masoni, J.~Matthews, U.-G. Meißner, I.-A.
  Melzer-Pellmann, M.~Mikhasenko, D.~J. Miller, D.~Milstead, R.~E. Mitchell,
  K.~Mönig, P.~Molaro, F.~Moortgat, M.~Moskovic, K.~Nakamura, M.~Narain,
  P.~Nason, S.~Navas, A.~Nelles, M.~Neubert, P.~Nevski, Y.~Nir, K.~A. Olive,
  C.~Patrignani, J.~A. Peacock, V.~A. Petrov, E.~Pianori, A.~Pich, A.~Piepke,
  F.~Pietropaolo, A.~Pomarol, S.~Pordes, S.~Profumo, A.~Quadt, K.~Rabbertz,
  J.~Rademacker, G.~Raffelt, M.~Ramsey-Musolf, B.~N. Ratcliff, P.~Richardson,
  A.~Ringwald, D.~J. Robinson, S.~Roesler, S.~Rolli, A.~Romaniouk, L.~J.
  Rosenberg, J.~L. Rosner, G.~Rybka, M.~G. Ryskin, R.~A. Ryutin, Y.~Sakai,
  S.~Sarkar, F.~Sauli, O.~Schneider, S.~Schönert, K.~Scholberg, A.~J.
  Schwartz, J.~Schwiening, D.~Scott, F.~Sefkow, U.~Seljak, V.~Sharma, S.~R.
  Sharpe, V.~Shiltsev, G.~Signorelli, M.~Silari, F.~Simon, T.~Sjöstrand,
  P.~Skands, T.~Skwarnicki, G.~F. Smoot, A.~Soffer, M.~S. Sozzi, S.~Spanier,
  C.~Spiering, A.~Stahl, S.~L. Stone, Y.~Sumino, M.~J. Syphers, F.~Takahashi,
  M.~Tanabashi, J.~Tanaka, M.~Taševský, K.~Terao, K.~Terashi, J.~Terning,
  R.~S. Thorne, M.~Titov, N.~P. Tkachenko, D.~R. Tovey, K.~Trabelsi,
  P.~Urquijo, G.~Valencia, R.~Van~de Water, N.~Varelas, G.~Venanzoni, L.~Verde,
  I.~Vivarelli, P.~Vogel, W.~Vogelsang, V.~Vorobyev, S.~P. Wakely,
  W.~Walkowiak, C.~W. Walter, D.~Wands, D.~H. Weinberg, E.~J. Weinberg,
  N.~Wermes, M.~White, L.~R. Wiencke, S.~Willocq, C.~G. Wohl, C.~L. Woody,
  W.-M. Yao, M.~Yokoyama, R.~Yoshida, G.~Zanderighi, G.~P. Zeller, O.~V. Zenin,
  R.-Y. Zhu, S.-L. Zhu, F.~Zimmermann, and P.~A. Zyla, ``{Review of Particle
  Physics},'' {\em Progress of Theoretical and Experimental Physics},
  vol.~2022, no.~8, p.~083C01, 2022.

\bibitem{bb1}
J.~E. Bjorkman and D.~R.~T. Jones, ``{The Unification Mass, $Sin^2\theta_W$ and
  $M_b / M_\tau$ in Nonminimal Supersymmetric SU(5)},'' {\em Nucl. Phys. B},
  vol.~259, p.~533, 1985.

\bibitem{qcd}
M.~Patgiri and N.~N. Singh, ``{New uncertainties in QCD-QED rescaling factors
  using quadrature method},'' {\em Pramana}, vol.~65, pp.~1015--1025, 2006.

\bibitem{g}
S.~Alam, M.~Aubert, S.~Avila, C.~Balland, J.~E. Bautista, M.~A. Bershady,
  D.~Bizyaev, M.~R. Blanton, A.~S. Bolton, J.~Bovy, J.~Brinkmann, J.~R.
  Brownstein, E.~Burtin, S.~Chabanier, M.~J. Chapman, P.~D. Choi, C.-H. Chuang,
  J.~Comparat, M.-C. Cousinou, A.~Cuceu, K.~S. Dawson, S.~de~la Torre,
  A.~de~Mattia, V.~d.~S. Agathe, H.~d.~M. des Bourboux, S.~Escoffier,
  T.~Etourneau, J.~Farr, A.~Font-Ribera, P.~M. Frinchaboy, S.~Fromenteau,
  H.~Gil-Mar\'{\i}n, J.-M. Le~Goff, A.~X. Gonzalez-Morales, V.~Gonzalez-Perez,
  K.~Grabowski, J.~Guy, A.~J. Hawken, J.~Hou, H.~Kong, J.~Parker, M.~Klaene,
  J.-P. Kneib, S.~Lin, D.~Long, B.~W. Lyke, A.~de~la Macorra, P.~Martini,
  K.~Masters, F.~G. Mohammad, J.~Moon, E.-M. Mueller, A.~Mu\~noz Guti\'errez,
  A.~D. Myers, S.~Nadathur, R.~Neveux, J.~A. Newman, P.~Noterdaeme, A.~Oravetz,
  D.~Oravetz, N.~Palanque-Delabrouille, K.~Pan, R.~Paviot, W.~J. Percival,
  I.~P\'erez-R\`afols, P.~Petitjean, M.~M. Pieri, A.~Prakash, A.~Raichoor,
  C.~Ravoux, M.~Rezaie, J.~Rich, A.~J. Ross, G.~Rossi, R.~Ruggeri,
  V.~Ruhlmann-Kleider, A.~G. S\'anchez, F.~J. S\'anchez, J.~R.
  S\'anchez-Gallego, C.~Sayres, D.~P. Schneider, H.-J. Seo, A.~Shafieloo,
  A.~c.~v. Slosar, A.~Smith, J.~Stermer, A.~Tamone, J.~L. Tinker, R.~Tojeiro,
  M.~Vargas-Maga\~na, A.~Variu, Y.~Wang, B.~A. Weaver, A.-M. Weijmans,
  C.~Y\`eche, P.~Zarrouk, C.~Zhao, G.-B. Zhao, and Z.~Zheng, ``Completed
  sdss-iv extended baryon oscillation spectroscopic survey: Cosmological
  implications from two decades of spectroscopic surveys at the apache point
  observatory,'' {\em Phys. Rev. D}, vol.~103, p.~083533, Apr 2021.

\bibitem{h}
N.~Aghanim, Y.~Akrami, M.~Ashdown, J.~Aumont, C.~Baccigalupi, M.~Ballardini,
  A.~J. Banday, R.~B. Barreiro, N.~Bartolo, S.~Basak, R.~Battye, K.~Benabed,
  J.-P. Bernard, M.~Bersanelli, P.~Bielewicz, J.~J. Bock, J.~R. Bond,
  J.~Borrill, F.~R. Bouchet, F.~Boulanger, M.~Bucher, C.~Burigana, R.~C.
  Butler, E.~Calabrese, J.-F. Cardoso, J.~Carron, A.~Challinor, H.~C. Chiang,
  J.~Chluba, L.~P.~L. Colombo, C.~Combet, D.~Contreras, B.~P. Crill,
  F.~Cuttaia, P.~de~Bernardis, G.~de~Zotti, J.~Delabrouille, J.-M. Delouis,
  E.~D. Valentino, J.~M. Diego, O.~Dor{\'{e} }, M.~Douspis, A.~Ducout,
  X.~Dupac, S.~Dusini, G.~Efstathiou, F.~Elsner, T.~A. En{\ss}lin, H.~K.
  Eriksen, Y.~Fantaye, M.~Farhang, J.~Fergusson, R.~Fernandez-Cobos,
  F.~Finelli, F.~Forastieri, M.~Frailis, A.~A. Fraisse, E.~Franceschi,
  A.~Frolov, S.~Galeotta, S.~Galli, K.~Ganga, R.~T. G{\'{e}}nova-Santos,
  M.~Gerbino, T.~Ghosh, J.~Gonz{\'{a}}lez-Nuevo, K.~M. G{\'{o}}rski,
  S.~Gratton, A.~Gruppuso, J.~E. Gudmundsson, J.~Hamann, W.~Handley, F.~K.
  Hansen, D.~Herranz, S.~R. Hildebrandt, E.~Hivon, Z.~Huang, A.~H. Jaffe, W.~C.
  Jones, A.~Karakci, E.~Keihänen, R.~Keskitalo, K.~Kiiveri, J.~Kim, T.~S.
  Kisner, L.~Knox, N.~Krachmalnicoff, M.~Kunz, H.~Kurki-Suonio, G.~Lagache,
  J.-M. Lamarre, A.~Lasenby, M.~Lattanzi, C.~R. Lawrence, M.~L. Jeune,
  P.~Lemos, J.~Lesgourgues, F.~Levrier, A.~Lewis, M.~Liguori, P.~B. Lilje,
  M.~Lilley, V.~Lindholm, M.~L{\'{o}}pez-Caniego, P.~M. Lubin, Y.-Z. Ma, J.~F.
  Mac{\'{\i}}as-P{\'{e}}rez, G.~Maggio, D.~Maino, N.~Mandolesi, A.~Mangilli,
  A.~Marcos-Caballero, M.~Maris, P.~G. Martin, M.~Martinelli,
  E.~Mart{\'{\i}}nez-Gonz{\'{a}}lez, S.~Matarrese, N.~Mauri, J.~D. McEwen,
  P.~R. Meinhold, A.~Melchiorri, A.~Mennella, M.~Migliaccio, M.~Millea,
  S.~Mitra, M.-A. Miville-Desch{\^{e}}nes, D.~Molinari, L.~Montier,
  G.~Morgante, A.~Moss, P.~Natoli, H.~U. N{\o}rgaard-Nielsen, L.~Pagano,
  D.~Paoletti, B.~Partridge, G.~Patanchon, H.~V. Peiris, F.~Perrotta,
  V.~Pettorino, F.~Piacentini, L.~Polastri, G.~Polenta, J.-L. Puget, J.~P.
  Rachen, M.~Reinecke, M.~Remazeilles, A.~Renzi, G.~Rocha, C.~Rosset,
  G.~Roudier, J.~A. Rubi{\~{n}}o-Mart{\'{\i}}n, B.~Ruiz-Granados, L.~Salvati,
  M.~Sandri, M.~Savelainen, D.~Scott, E.~P.~S. Shellard, C.~Sirignano,
  G.~Sirri, L.~D. Spencer, R.~Sunyaev, A.-S. Suur-Uski, J.~A. Tauber,
  D.~Tavagnacco, M.~Tenti, L.~Toffolatti, M.~Tomasi, T.~Trombetti,
  L.~Valenziano, J.~Valiviita, B.~V. Tent, L.~Vibert, P.~Vielva, F.~Villa,
  N.~Vittorio, B.~D. Wandelt, I.~K. Wehus, M.~White, S.~D.~M. White,
  A.~Zacchei, and A.~Zonca, ``{Planck 2018 results. VI. Cosmological
  parameters},'' {\em Astron. Astrophys.}, vol.~641, p.~A6, 2020.
\newblock [Erratum: Astron.Astrophys. 652, C4 (2021)].

\bibitem{de}
P.~F. de~Salas, D.~V. Forero, S.~Gariazzo, P.~Mart\'\i{}nez-Mirav\'e, O.~Mena,
  C.~A. Ternes, M.~T\'ortola, and J.~W.~F. Valle, ``{2020 global reassessment
  of the neutrino oscillation picture},'' {\em JHEP}, vol.~02, p.~071, 2021.

\bibitem{higgs}
A.~Sirunyan, A.~Tumasyan, and other, ``A measurement of the higgs boson mass in
  the diphoton decay channel,'' {\em Physics Letters B}, vol.~805, p.~135425,
  2020.

\bibitem{rg2}
S.~Antusch, J.~Kersten, M.~Lindner, and M.~Ratz, ``{Running neutrino masses,
  mixings and CP phases: Analytical results and phenomenological
  consequences},'' {\em Nucl. Phys. B}, vol.~674, pp.~401--433, 2003.

\end{thebibliography}

\end{document}